\documentclass[fleqn,usenatbib]{mnras}

\usepackage{newtxtext,newtxmath}

\usepackage[T1]{fontenc}
\usepackage{ae,aecompl}
\usepackage{natbib}

\def\fesc{\ifmmode f_{\rm esc} \else $f_{\rm esc}$\fi}

\usepackage{graphicx}	
\usepackage{amsmath}	
\usepackage{amssymb}	

\title[Ly$\alpha$ in galaxies with high O$_{32}$ ratios]{Diverse properties
of Ly$\alpha$ emission in low-redshift compact star-forming galaxies with 
extremely high [O~{\sc iii}]/[O~{\sc ii}] ratios}

\author[Y. I. Izotov et al.]{
Y. I. Izotov$^{1,2}$, 
D. Schaerer$^{3,4}$, 
G. Worseck$^{5}$, 
A. Verhamme$^{3}$,
N. G. Guseva$^{1,2}$, 
\newauthor 
~T. X. Thuan$^{6}$, I. Orlitov\'a$^{7}$ \& K. J. Fricke$^{8,2}$
\\
$^{1}$Bogolyubov Institute for Theoretical Physics,
National Academy of Sciences of Ukraine, 14-b Metrolohichna str., Kyiv,
03143, Ukraine,\\
E-mail: yizotov@bitp.kiev.ua, nguseva@bitp.kiev.ua\\
$^{2}$Max-Planck-Institut f\"ur Radioastronomie, Auf dem H\"ugel 
69, 53121 Bonn, Germany\\
$^{3}$Observatoire de Gen\`eve, Universit\'e de Gen\`eve, 
51 Ch. des Maillettes, 1290, Versoix, Switzerland,\\
E-mail: daniel.schaerer@unige.ch, anne.verhamme@unige.ch\\
$^{4}$IRAP/CNRS, 14, Av. E. Belin, 31400 Toulouse, France\\
$^{5}$ Institut f\"ur Physik und Astronomie, Universit\"at Potsdam, Karl-Liebknecht-Str. 24/25, D-14476 Potsdam, Germany,\\
E-mail: gworseck@uni-potsdam.de\\
$^{6}$Astronomy Department, University of Virginia, P.O. Box 400325, 
Charlottesville, VA 22904-4325, USA,\\
E-mail: txt@virginia.edu\\
$^{7}$Astronomical Institute, Czech Academy of Sciences, Bo\v cn{\'\i} II 1401, 
141 00, Prague, Czech Republic,\\
E-mail: orlitova@asu.cas.cz\\
$^{8}$Institut f\"ur Astrophysik, G\"ottingen Universit\"at, 
Friedrich-Hund-Platz 1, 37077 G\"ottingen, Germany,\\
E-mail: kfricke@gwdg.de}

\date{Accepted XXX. Received YYY; in original form ZZZ}

\pubyear{2019}

\begin{document}
\label{firstpage}
\pagerange{\pageref{firstpage}--\pageref{lastpage}}
\maketitle

\begin{abstract}
We present observations with the Cosmic 
Origins Spectrograph onboard the {\sl Hubble Space Telescope} of
eight compact star-forming galaxies at redshifts $z$ = 0.02811 -- 0.06540, 
with low oxygen abundances 12 + log(O/H) = 7.43 --  7.82 and 
extremely high emission-line flux ratios 
O$_{32}$ = [O~{\sc iii}]$\lambda$5007/[O~{\sc ii}]$\lambda$3727 $\sim$ 22 -- 39,
aiming to study the properties of Ly$\alpha$ emission in such conditions. 
We find a diversity in Ly$\alpha$ properties. In five galaxies 
Ly$\alpha$ emission line is strong, with equivalent width (EW) in the range 
45 -- 190\AA. In the remaining galaxies, weak Ly$\alpha$ emission with 
EW(Ly$\alpha$)
$\sim$ 2 -- 7\AA\ is superposed on a broad Ly$\alpha$ absorption line,
indicating a high neutral hydrogen column density 
$N$(H~{\sc i})~$\sim$~(1~--~3)$\times$10$^{21}$ cm$^{-2}$.
We examine the relation between the Ly$\alpha$ escape fraction 
$f_{\rm esc}$(Ly$\alpha$) and the Lyman continuum escape fraction 
$f_{\rm esc}$(LyC), using direct measures of the latter in eleven low-redshift
LyC leakers, to verify whether $f_{\rm esc}$(Ly$\alpha$) can be an
indirect measure of escaping LyC radiation. 
The usefulness of O$_{32}$, of the Ly$\alpha$ equivalent width 
EW(Ly$\alpha$) and of the Ly$\alpha$ peak separation $V_{\rm sep}$
as indirect indicators of Ly$\alpha$ leakage is also discussed.
It is shown that there is no correlation between O$_{32}$ and 
$f_{\rm esc}$(Ly$\alpha$). We find an increase of 
$f_{\rm esc}$(Ly$\alpha$) with increasing EW(Ly$\alpha$) for  
EW(Ly$\alpha$)$<$100\AA, but for higher EW(Ly$\alpha$)$\ga$150\AA\  
the $f_{\rm esc}$(Ly$\alpha$) is nearly constant attaining the value of
$\sim$0.25. We find an anticorrelation between $f_{\rm esc}$(Ly$\alpha$) and
$V_{\rm sep}$, though not as tight as the one found earlier between 
$f_{\rm esc}$(LyC) and $V_{\rm sep}$. This finding makes $V_{\rm sep}$ 
a promising indirect indicator of both
the Ly$\alpha$ and ionizing radiation leakage.
\end{abstract}




\section{Introduction}\label{intro}

The extremely high O$_{32}$ = 
[O~{\sc iii}]$\lambda$5007/[O~{\sc ii}]$\lambda$3727 ratio
in spectra of low-redshift dwarf star-forming galaxies (SFGs) indicates the 
presence of either an ionization-bounded compact H~{\sc ii} region with
intense ionizing radiation from a very young star-forming region or 
a density-bounded H~{\sc ii} region
losing its ionizing radiation, or both \citep[e.g. ][]{JO13,NO14,S15,I17}. 
The ionizing photon production efficiency $\xi_{\rm ion}$~=~$N_{\rm LyC}$/$L_\nu$
in these galaxies could be very high, where $N_{\rm LyC}$ is the Lyman continuum 
photon production rate and $L_\nu$ is the monochromatic UV luminosity
at $\lambda$ = 1500\AA. 
Therefore, they are considered as the local counterparts of 
the dwarf SFGs at high redshifts, which are 
likely the main source of the last phase transition of the matter in the 
Universe, namely its reionization, which ends at $z$~$\sim$~6 
\citep*{O09,WC09,K11,J13,M13,Y11,B15a,Sm18,N18,St18,K19}. 
An alternative hypothesis is the dominant role of active galactic nuclei (AGN) 
in the reionization of the Universe, as proposed e.g. by \citet{Madau15}. However,
in recent years, growing evidence for a small contribution of AGN to reionization
has accumulated \citep*[e.g. ][]{H18,M18,P18,Mat18,Ku19}.
On the other hand, SFGs can be considered as an important source of ionization
only if the product $\xi_{\rm ion}$$\times$$f_{\rm esc}$(LyC) is sufficiently high, where
$f_{\rm esc}$(LyC) is the escape fraction of Lyman continuum (LyC), i.e. the
fraction of ionizing radiation emitted into the intergalactic medium (IGM). 
It has been estimated that
$f_{\rm esc}$(LyC) should not be less than 10--20 per cent on average in order 
to accomplish reionization \citep[e.g.][]{O09,R13,D15,Robertson15,K16}.

Currently, direct observations of the LyC are available for only a limited number of
galaxies at high and low redshifts. Searches for and confirmation of LyC
emission from high redshift galaxies
are difficult due to their faintness, contamination by lower-redshift
interlopers and attenuation by residual H~{\sc i} in the highly ionized IGM.
\citep[e.g., ][]{V10,V12,Inoue14,Gr15}. They have  
resulted in the discovery of approximately three dozens of
LyC leakers with reliably high escape fraction: {\em Ion2} \citep{Va15,B16},
Q1549-C25 \citep{Sh16}, A2218-Flanking \citep{B17}, {\em Ion3} \citep{Va18}, 
SunBurst Arc \citep{RT19}, 15 SFGs studied by \citet{St18}, and a dozen of
galaxies at $z$ = 3.1 with strong [O~{\sc iii}] $\lambda$5007 emission
\citep{F19}.

At low redshifts, $z$ $\sim$ 0.3 - 0.4, \citet{I16,I16b,I18,I18b}
have detected LyC leaking emission in eleven compact SFGs with high 
O$_{32}$ $\sim$ 5 - 28 and derived $f_{\rm esc}$(LyC) ranging 
between 2 and 72 per cent. 
These values are higher than $f_{\rm esc}$(LyC) of $\la$ 1 per cent 
derived  by \citet{B14} and \citet{C17} in SFGs with lower O$_{32}$, or by 
\citet{He18} in SFGs with lower equivalent widths of the H$\beta$ emission line.
There is hope that the number of low-$z$ LyC leakers will considerably
be increased after the completion in 2020 of the {\sl HST} Large Program 
GO 15626 (P.I. A. Jaskot), which includes more than 70 targets.

The {\sl Hubble Space Telescope} ({\sl HST})/Cosmic Origins Spectrograph (COS) 
sensitivity curve limits LyC observations to $z \ga 0.25$. At lower 
redshifts, only very bright targets are accessible \citep[e.g. ][]{L16}.
Therefore, for low-redshift SFGs,
reliable indirect indicators of LyC leakage are needed. \citet{NO14} proposed
that high O$_{32}$ can be an indication of density-bounded H~{\sc ii} region and
thus of LyC leakage. However, O$_{32}$ depends on some other parameters, such as
ionization parameter and hardness of radiation which in turn depend on 
metallicity.

\citet{F16} has found a correlation between $f_{\rm esc}$(LyC) and O$_{32}$, 
using a small sample of galaxies which includes low-$z$ LyC leakers with
O$_{32}$ $<$ 8. On the other hand, no correlation has been found by 
\citet{I18b} for eleven confirmed LyC 
leakers with O$_{32}$ in a wide range of $\sim$ 5 - 28. A similar conclusion
was obtained by \citet{N18} and \citet{B19} \citep[see also ][]{R17}. 
However, this conclusion is based 
mainly on observations of galaxies with O$_{32}$ $\la$ 16. Only one observed
galaxy has a higher value of O$_{32}$. It is thus important to 
observe LyC emission in more galaxies with higher O$_{32}$.

The He~{\sc i} emission-line ratios in optical spectra of galaxies can also be 
used to estimate the column density of neutral gas and thus to indirectly
estimate $f_{\rm esc}$(LyC) \citep{I17}. However, deep spectra in the 
optical range are needed since these lines are weak with intensities $\la$10
per cent of the H$\beta$ emission line.

Finally, the properties of the Ly$\alpha$ emission line can serve as indirect 
indicators of escaping LyC emission. Its profile is double-peaked in
the spectra of most of LyC leakers, with the separation
between the peaks $V_{\rm sep}$ decreasing with decreasing neutral hydrogen
column density $N$(H~{\sc i}) \citep[e.g. ][]{I16,I16b,I18,I18b}. 
In rare cases, the Ly$\alpha$ profile is more complex and consists of three and
more peaks \citep{RT17,I18b,Va18}.
\citet{V15} and \citet{I18b} have shown that
there is a tight relation between the LyC escape fraction $f_{\rm esc}$(LyC) and
the separation between Ly$\alpha$ peaks, making $V_{\rm sep}$ the most reliable 
indicator of escaping ionizing radiation. Furthermore, \citet{V17} have shown
for a limited sample of LyC leakers that the Ly$\alpha$ escape fractions
$f_{\rm esc}$(Ly$\alpha$) are larger than the LyC escape fractions 
$f_{\rm esc}$(LyC), in accord with theoretical predictions of \citet*{D16}.

  \begin{table*}
  \caption{Some general characteristics of the selected galaxies from the SDSS data base
\label{tab1}}
\begin{tabular}{lrrccccccc} \hline
Name&R.A.(2000.0)&Dec.(2000.0)&$z$&$D_L$$^{\rm a}$&$D_A$$^{\rm b}$&O$_{32}$$^{\rm c}$
&EW(H$\beta$)&FWHM$^{\rm d}$&12+log(O/H)$^{\rm e}$ \\ 
    &            &            &   & (Mpc)       &  (Mpc)      &
&(\AA)&(arcsec)& \\ \hline
J0007$+$0226&00:07:24.49&$+$02:26:27.20&0.06360&287& 254&35&258&1.18&7.82\\ 
J0159$+$0751&01:59:52.75&$+$07:51:48.80&0.06105&275& 245&39&322&1.05&7.56\\ 
J0820$+$5431&08:20:19.27&$+$54:31:40.10&0.03851&171& 158&22&334&1.29&7.49\\ 
J0926$+$4504&09:26:55.44&$+$45:04:32.30&0.04232&188& 173&22&328&1.42&7.68\\ 
J1032$+$4919&10:32:56.72&$+$49:19:47.24&0.04420&197& 181&24&444&1.27&7.59\\ 
J1205$+$4551&12:05:03.55&$+$45:51:50.94&0.06540&296& 261&23&483&1.37&7.44\\ 
J1242$+$4851&12:42:26.46&$+$48:51:57.67&0.06226&281& 249&25&360&1.25&7.43\\ 
J1355$+$4651&13:55:25.66&$+$46:51:51.34&0.02811&124& 117&23&269&1.35&7.56\\ 
\hline
\end{tabular}

\hbox{$^{\rm a}$Luminosity distance \citep[NED, ][]{W06}.}

\hbox{$^{\rm b}$Angular size distance \citep[NED, ][]{W06}.}

\hbox{$^{\rm c}$O$_{32}$ = [O~{\sc iii}]$\lambda$5007/[O~{\sc ii}]$\lambda$3727.
For J1032$+$4919, O$_{32}$ = 3$\times$[O~{\sc iii}]$\lambda$4959/[O~{\sc ii}]$\lambda$3727.}

\hbox{$^{\rm d}$Full width at half maximum in the SDSS $g$-band image.}

\hbox{$^{\rm e}$Oxygen abundance derived from line intensities in the SDSS spectrum.}

  \end{table*}

  \begin{table*}
  \caption{Apparent magnitudes with errors in parentheses compiled
from the SDSS, {\sl GALEX}$^{\rm a}$ and {\sl WISE}$^{\rm b}$ databases
\label{tab2}}
\begin{tabular}{lccccccccccccc} \hline
Name&\multicolumn{5}{c}{SDSS (AB)}
&&\multicolumn{2}{c}{{\sl GALEX} (AB)}&&\multicolumn{4}{c}{{\sl WISE} (Vega)} \\ 
    &\multicolumn{1}{c}{$u$}&\multicolumn{1}{c}{$g$}&\multicolumn{1}{c}{$r$}&\multicolumn{1}{c}{$i$}&\multicolumn{1}{c}{$z$}&&FUV&NUV&&\multicolumn{1}{c}{$W1$}&\multicolumn{1}{c}{$W2$}&\multicolumn{1}{c}{$W3$}&\multicolumn{1}{c}{$W4$}
\\
    &\multicolumn{1}{c}{(err)}&\multicolumn{1}{c}{(err)}&\multicolumn{1}{c}{(err)}&\multicolumn{1}{c}{(err)}&\multicolumn{1}{c}{(err)}&&(err)&(err)&&\multicolumn{1}{c}{(err)}&\multicolumn{1}{c}{(err)}&\multicolumn{1}{c}{(err)}&\multicolumn{1}{c}{(err)} \\
\hline
J0007$+$0226& 21.06& 19.73& 20.66& 19.82& 20.44&& 21.80& 21.28&&\multicolumn{1}{c}{17.40}&\multicolumn{1}{c}{17.03}&\multicolumn{1}{c}{11.92}&\multicolumn{1}{c}{ ... } \\
            &(0.13)&(0.02)&(0.05)&(0.03)&(0.23)&&(0.31)&(0.22)&&\multicolumn{1}{c}{(0.19)}&\multicolumn{1}{c}{(0.50)}&\multicolumn{1}{c}{(0.32)}&\multicolumn{1}{c}{(...)} \\
J0159$+$0751& 20.67& 19.47& 20.36& 19.57& 19.89&& 21.21& 20.43&&\multicolumn{1}{c}{ 17.39}&\multicolumn{1}{c}{ 15.88}&\multicolumn{1}{c}{ 11.54}&\multicolumn{1}{c}{ 8.52} \\
            &(0.11)&(0.02)&(0.05)&(0.04)&(0.18)&&(0.27)&(0.29)&&\multicolumn{1}{c}{(0.15)}&\multicolumn{1}{c}{(0.15)}&\multicolumn{1}{c}{(0.19)}&\multicolumn{1}{c}{(0.30)} \\
J0820$+$5431& 21.44& 20.41& 21.55& 21.44& 21.47&& 21.67& 21.59&&\multicolumn{1}{c}{16.52}&\multicolumn{1}{c}{16.59}&\multicolumn{1}{c}{ ... }&\multicolumn{1}{c}{ ... } \\
            &(0.13)&(0.02)&(0.09)&(0.12)&(0.51)&&(0.34)&(0.24)&&\multicolumn{1}{c}{(0.07)}&\multicolumn{1}{c}{(0.25)}&\multicolumn{1}{c}{(...)}&\multicolumn{1}{c}{(...)} \\
J0926$+$4504& 21.10& 19.92& 21.11& 20.68& 20.54&& 22.12& 21.74&&\multicolumn{1}{c}{ ... }&\multicolumn{1}{c}{17.26}&\multicolumn{1}{c}{12.37}&\multicolumn{1}{c}{ ... } \\
            &(0.11)&(0.02)&(0.08)&(0.07)&(0.23)&&(0.44)&(0.32)&&\multicolumn{1}{c}{(...)}&\multicolumn{1}{c}{(0.48)}&\multicolumn{1}{c}{(0.44)}&\multicolumn{1}{c}{(...)} \\
J1032$+$4919& 19.89& 18.67& 20.01& 19.60& 20.15&& 20.26& 20.25&& 16.60& 15.08& 10.50&\multicolumn{1}{c}{7.74} \\
            &(0.04)&(0.01)&(0.02)&(0.02)&(0.15)&&(0.20)&(0.14)&&\multicolumn{1}{c}{(0.08)}&\multicolumn{1}{c}{(0.07)}&\multicolumn{1}{c}{(0.07)}&\multicolumn{1}{c}{(14)} \\
J1205$+$4551& 20.52& 19.79& 20.98& 19.61& 20.80&& 21.04& 20.58&&\multicolumn{1}{c}{ 15.14}&\multicolumn{1}{c}{ 13.55}&\multicolumn{1}{c}{ 9.91}&\multicolumn{1}{c}{ 7.80} \\
            &(0.05)&(0.01)&(0.04)&(0.02)&(0.16)&&(0.34)&(0.19)&&\multicolumn{1}{c}{(0.04)}&\multicolumn{1}{c}{(0.03)}&\multicolumn{1}{c}{(0.05)}&\multicolumn{1}{c}{(0.19)} \\
J1242$+$4851& 21.30& 20.35& 21.51& 20.56& 21.25&& 21.36& 21.36&&\multicolumn{1}{c}{18.37}&\multicolumn{1}{c}{ ...}&\multicolumn{1}{c}{ ...}&\multicolumn{1}{c}{ ...} \\
            &(0.09)&(0.02)&(0.07)&(0.05)&(0.31)&&(0.17)&(0.14)&&\multicolumn{1}{c}{(0.33)}&\multicolumn{1}{c}{(...)}&\multicolumn{1}{c}{(...)}&\multicolumn{1}{c}{(...)} \\
J1355$+$4651& 20.37& 19.22& 20.06& 20.47& 20.53&& 20.54& 20.66&&\multicolumn{1}{c}{17.86}&\multicolumn{1}{c}{17.39}&\multicolumn{1}{c}{ ...}&\multicolumn{1}{c}{ ...} \\
            &(0.05)&(0.01)&(0.02)&(0.05)&(0.20)&&(0.23)&(0.19)&&\multicolumn{1}{c}{(0.19)}&\multicolumn{1}{c}{(0.47)}&\multicolumn{1}{c}{(...)}&\multicolumn{1}{c}{(...)} \\
\hline
\end{tabular}

\hbox{$^{\rm a}$http://galex.stsci.edu/GR6/}

\hbox{$^{\rm b}$https://irsa.ipac.caltech.edu/cgi-bin/Gator/nph-dd}

  \end{table*}

\begin{figure*}
\hbox{
\includegraphics[angle=0,width=0.24\linewidth]{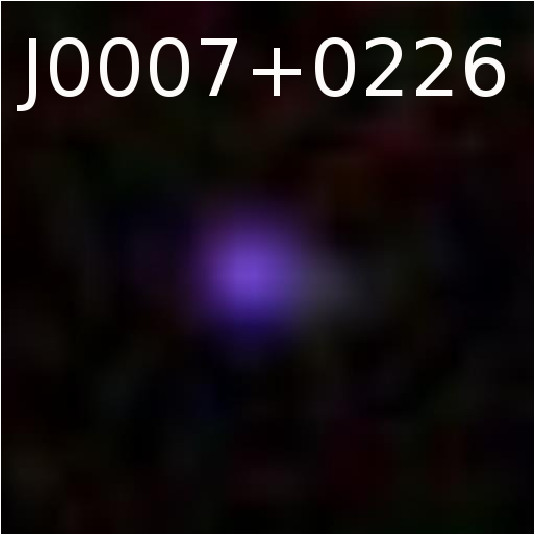}
\includegraphics[angle=0,width=0.24\linewidth]{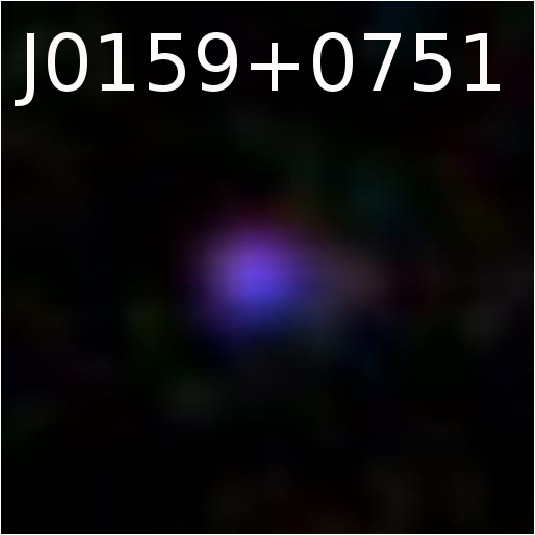}
\includegraphics[angle=0,width=0.24\linewidth]{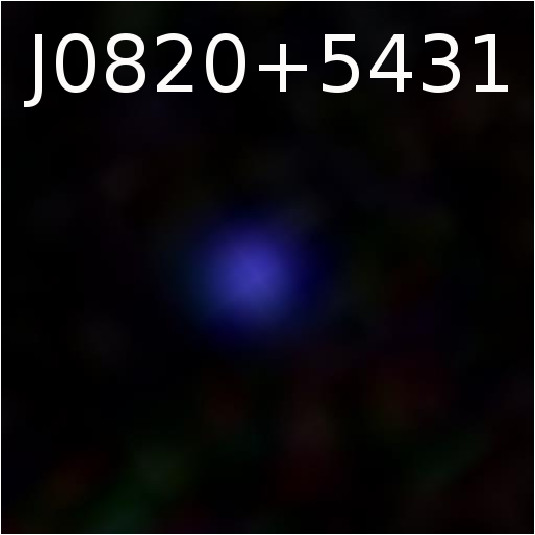}
\includegraphics[angle=0,width=0.24\linewidth]{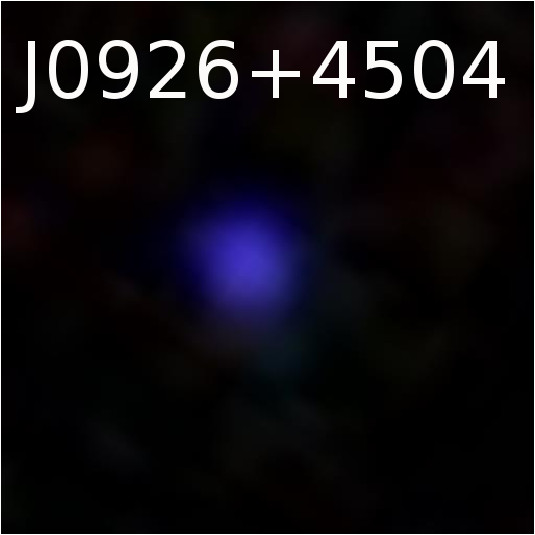}
}
\hbox{
\includegraphics[angle=0,width=0.24\linewidth]{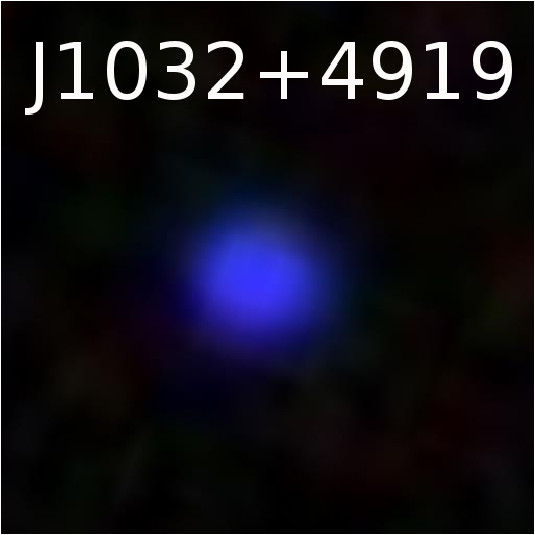}
\includegraphics[angle=0,width=0.24\linewidth]{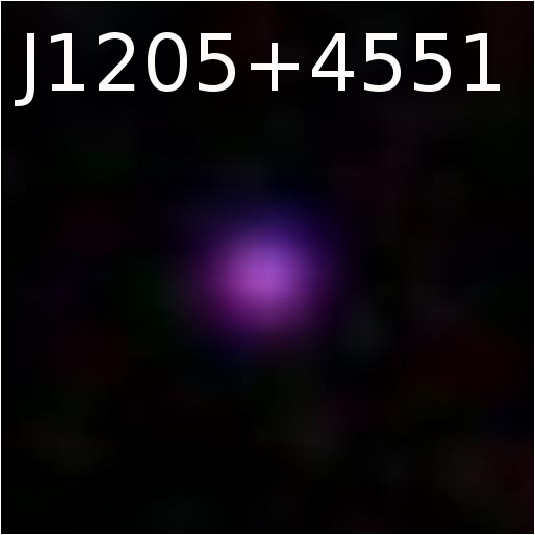}
\includegraphics[angle=0,width=0.24\linewidth]{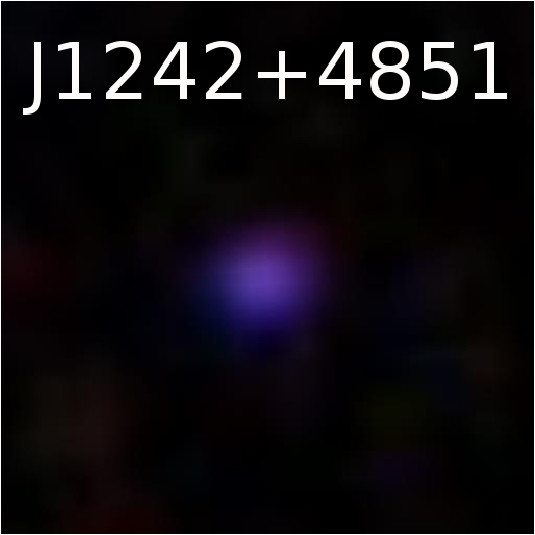}
\includegraphics[angle=0,width=0.24\linewidth]{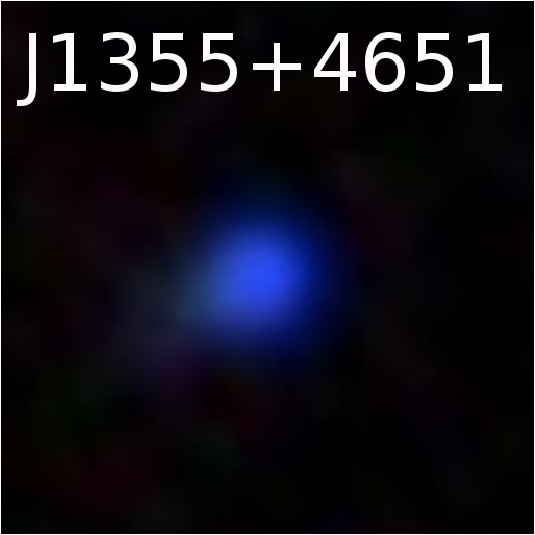}
}
\caption{12 arcsec $\times$ 12 arcsec SDSS composite images of galaxies
with extreme O$_{32}$ ratios observed with the {\sl HST}.
\label{fig1}}
\end{figure*}

\begin{figure*}
\includegraphics[angle=-90,width=0.49\linewidth]{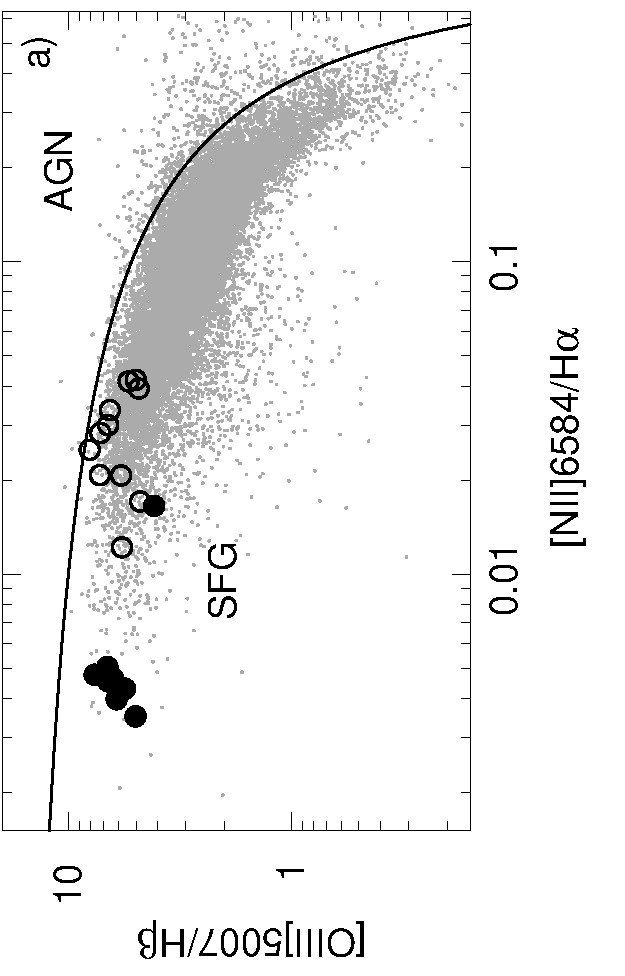}
\includegraphics[angle=-90,width=0.49\linewidth]{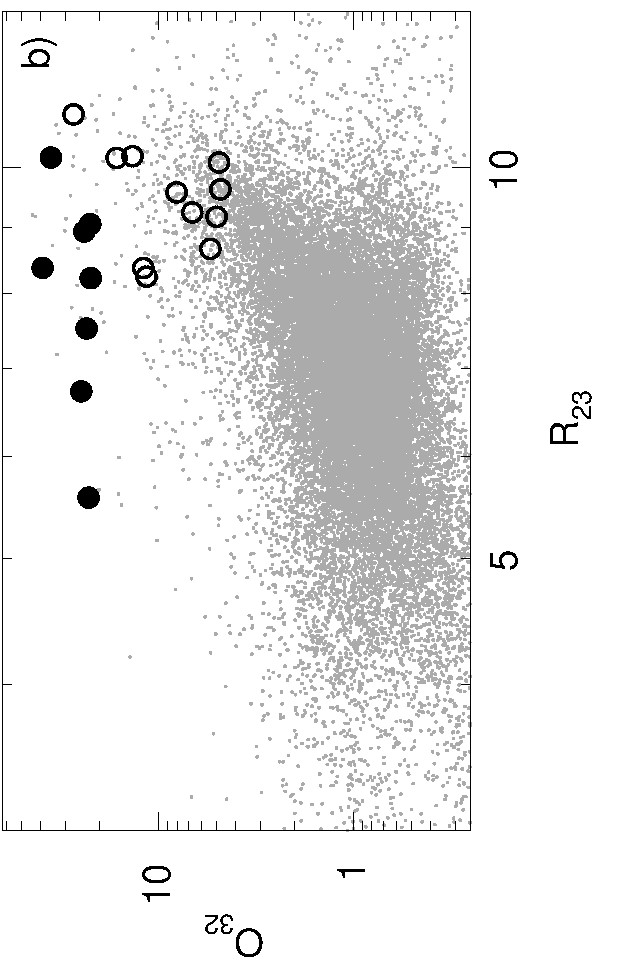}
\caption{{\bf a)} The Baldwin-Phillips-Terlevich (BPT) diagram \citep{BPT81} for
SFGs. {\bf b)} The O$_{32}$ -- R$_{23}$ diagram for SFGs where R$_{23}$ = 
([O~{\sc ii}]3727 + [O~{\sc iii}]4959 + [O~{\sc iii}]5007)/H$\beta$.
In both panels, the galaxies  from this paper and LyC leaking galaxies 
\citep{I16,I16b,I18,I18b} are shown by filled circles and open circles, 
respectively.
The compact SFGs from the SDSS \citep{I16c} are represented by grey dots.
The solid line in {\bf a)} by \citet{K03} separates SFGs from active galactic 
nuclei (AGN).
\label{fig2}}
\end{figure*}

In this paper we present new {\sl HST}/COS observations of the Ly$\alpha$
in eight low-redshift ($z$ $<$ 0.07) compact SFGs with the highest 
O$_{32}$ $\sim$ 22 -- 39 found in the Sloan Digital Sky Survey (SDSS),
aiming to examine the properties of Ly$\alpha$ emission in these galaxies and to
verify possible indirect indicators of escaping Ly$\alpha$ and LyC emission. 
The studied galaxies have lower stellar masses than previously observed 
Ly$\alpha$ emitting galaxies such as the ``green pea'' (GP) galaxies at 
$z$ $\sim$ 0.1 -- 0.3
studied e.g. by \citet{J17}, \citet{Y17}, \citet{MK19}, and the confirmed LyC 
leakers at $z$ $\sim$ 0.3 -- 0.4 \citep{I16,I16b,I18,I18b}.
The selection criteria and the properties of the selected galaxies derived from 
observations in the optical range are presented in Section~\ref{sec:select}. 
The {\sl HST} observations and data reduction are described
in Section~\ref{sec:obs}. The surface brightness profiles in the UV range are 
discussed in Section~\ref{sec:sbp}. Ly$\alpha$ emission is considered in 
Section~\ref{sec:lya}.  In Section~\ref{discussion}, our results are compared 
with the Ly$\alpha$ escape fractions obtained for other galaxies in some recent 
studies. We summarize our findings in Section~\ref{summary}.

\section{Selection criteria and the integrated properties of selected galaxies 
}\label{sec:select}

Our galaxies were selected from the SDSS Data Release 14 (DR14) 
\citep{A18}. We adopted the following selection criteria for each galaxy:

1) it has a very high O$_{32}$ $>$ 20,
which is an indication of either a density-bounded H~{\sc ii} region or
a high ionization parameter.

2) it has a compact structure in the SDSS image in order to observe the total flux of the 
galaxy with the small 2.5 arcsec COS aperture.

3) it is isolated from other sources to avoid
contamination in the {\sl HST}/COS spectra.

4) it is sufficiently bright in the UV range ({\sl GALEX} NUV magnitude brighter
than 22 mag) to be acquired with COS.

In total, we selected the 8 brightest galaxies satisfying the above conditions.

The SDSS spectra of the selected galaxies reveal considerably lower oxygen 
abundances, in the range 12 + log(O/H) = 7.43 -- 7.82 (see below), as compared to the
confirmed LyC leakers \citep{I16,I16b,I18,I18b}. They show comparably 
high rest-frame equivalent widths EW(H$\beta$) = 258 -- 483~\AA\ 
(Table \ref{tab1}), indicating very recent star formation.

These galaxies are located at the extreme upper tip of the SFG branch in the 
Baldwin-Phillips-Terlevich (BPT) diagram (filled circles in 
Fig. \ref{fig2}a) implying the presence of high-excitation H~{\sc ii} regions.
They are considerably more extreme than the confirmed
LyC leakers shown by open circles.
It is seen in Fig.~\ref{fig2}b, which shows the 
O$_{32}$ -- R$_{23}$ diagram, that the selected SFGs (filled circles) have the highest O$_{32}$ ratios among all SDSS SFGs,  
and that they are offset to low R$_{23}$ values compared to the main
sequence of SFGs (grey dots) because of their lower metallicities. On the other
hand, most of the confirmed LyC leakers with higher metallicities are located 
in the upper part of the main sequence (open circles).

The SDSS spectra were used to derive the interstellar extinction from
the hydrogen Balmer decrement and ionized gas metallicity using the 
prescriptions of \citet*{ITL94}, \citet{I06} and \citet{G13}.

The extinction-corrected emission-line flux densities, the extinction 
coefficients $C$(H$\beta$), the rest-frame equivalent 
widths of H$\beta$, [O~{\sc iii}]$\lambda$5007 and H$\alpha$ emission lines, 
and the observed H$\beta$ flux densities are shown in Table \ref{taba1}.
We note that the H$\alpha$ emission line in the SDSS spectra of J0820$+$5431 
and J1032$+$4919 and the [O~{\sc iii}]$\lambda$5007 emission line in the SDSS
spectrum of J1032$+$4919 are clipped. Therefore, for J1032$+$4919,
$I$([O~{\sc iii}]$\lambda$5007) = 3$\times$$I$([O~{\sc iii}]$\lambda$4959) 
is used to derive O$_{32}$, the electron temperature $T_{\rm e}$(O~{\sc iii})
and oxygen abundance. Furthermore, the H$\alpha$ emission line in this
galaxy and in J0820$+$5431 is excluded in the calculation of extinction.
The He~{\sc ii} $\lambda$4686\AA\ emission line is detected in seven
out of eight galaxies implying hard ionizing radiation.

\begin{figure*}
\includegraphics[angle=-90,width=0.99\linewidth]{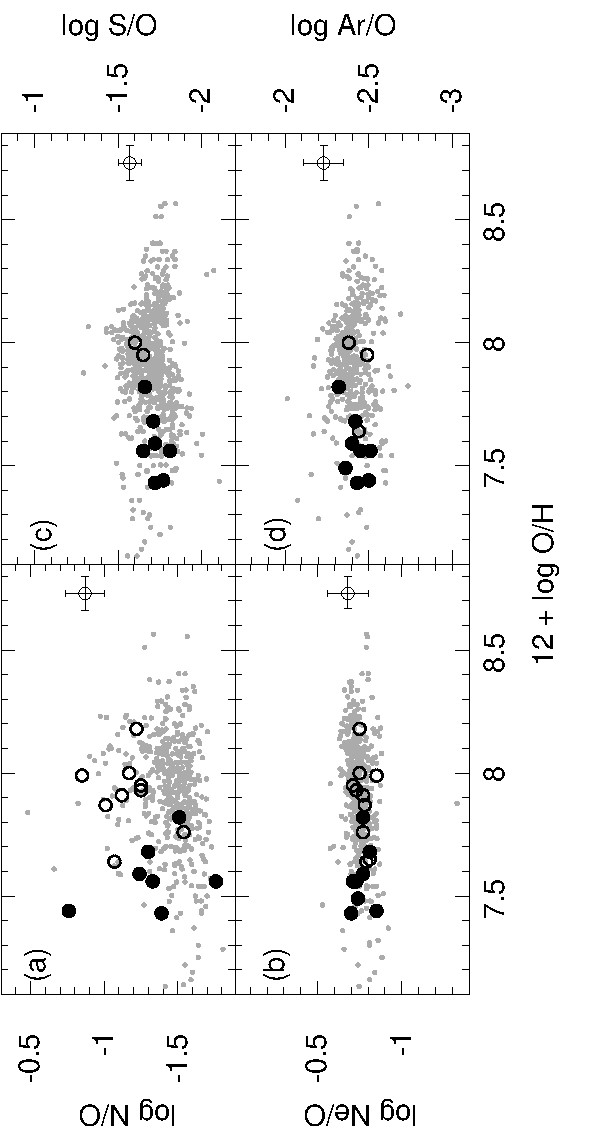}
\caption{Dependences of heavy element-to-oxygen abundance ratios on oxygen 
abundance. In all panels, galaxies from this paper and LyC leaking 
galaxies \citep{I16,I16b,I18,I18b} are shown by filled circles and open 
circles, respectively.
The compact SFGs from the SDSS \citep{I16c} are represented by grey dots.
Solar values by \citet{L03} are shown by open circles with error bars.
\label{fig3}}
\end{figure*}

The flux densities from Table \ref{taba1} and the direct $T_{\rm e}$ method are 
used to derive the electron temperature, electron number 
density and the element abundances in the H~{\sc ii} regions.
These quantities are shown in Table \ref{taba2}. The oxygen abundances 
of our galaxies are on average $\sim$ 0.2 - 0.3 dex lower than
those in known low-redshift LyC leakers 
\citep{I16,I16b,I18,I18b}. Four galaxies from our sample have been observed with 
the Large Binocular Telescope (LBT) by \citet{I17}. Those authors derived oxygen 
abundances that are in good agreement with the present determinations from SDSS 
spectra. The neon, sulfur and argon to oxygen abundance ratios (Fig.~\ref{fig3}) 
are similar to those in dwarf emission-line galaxies \citep[e.g. ][]{I06,G13} and
in confirmed LyC leakers, with similar small dispersions. On the other hand,
the N/O ratios are scattered over a large range (Fig.~\ref{fig3}a) which
can not be explained by N/O errors (Table \ref{taba2}). Compared to SDSS SFGs 
(shown by grey dots), they tend to have systematically higher values both for 
the galaxies in our sample and for the confirmed LyC leakers.

  \begin{table*}
  \caption{Integrated characteristics \label{tab3}}
  \begin{tabular}{lcccccccccccc} \hline
Name&$M_{\rm FUV}^{\rm SED}$$^{\rm a}$&$M_{\rm FUV}$$^{\rm b}$&  $M_g$$^{\rm c}$   &log $M_\star$$^{\rm d}$ &SB age$^{\rm e}$&log $L$(H$\beta$)$^{\rm f}$&SFR$^{\rm g}$&log $\xi_{\rm ion}$$^{\rm h}$&$\alpha$$^{\rm i}$&$r_{50}$$^{\rm j}$&$\Sigma_1$$^{\rm k}$&$\Sigma_2$$^{\rm l}$\\
\hline   
J0007$+$0226&$-$17.35&$-$15.49&$-$17.62&7.9&1.3&40.71&1.4&25.44&0.48&0.070&~\,1.9&~\,90\\
J0159$+$0751&$-$17.16&$-$15.99&$-$18.02&8.6&0.5&40.92&2.3&25.76&0.17&0.060&25.3&203\\
J0820$+$5431&$-$15.68&$-$14.49&$-$15.75&6.0&3.1&39.85&0.2&25.24&0.23&0.020&~\,1.2&159\\
J0926$+$4504&$-$15.44&$-$14.22&$-$16.51&7.2&2.8&39.91&0.2&25.41&0.23&0.020&~\,1.2&159\\
J1032$+$4919&$-$17.63&$-$16.21&$-$17.82&6.8&2.7&40.95&2.4&25.32&0.28&0.040&~\,9.7&478\\
J1205$+$4551&$-$17.67&$-$16.32&$-$17.63&6.9&2.4&41.08&2.7&25.69&0.30&0.060&~\,9.6&239\\
J1242$+$4851&$-$16.48&$-$15.88&$-$16.93&7.6&2.0&40.54&0.8&25.62& ...&...&...&...\\
J1355$+$4651&$-$14.88&$-$14.93&$-$16.31&5.8&3.3&39.68&0.1&25.37&0.09&0.016&~\,3.9&124\\
\hline
  \end{tabular}

\hbox{$^{\rm a}$$M_{\rm FUV}^{\rm SED}$ is the absolute FUV magnitude derived from the intrinsic rest-frame SED.}

\hbox{$^{\rm b}$$M_{\rm FUV}$ is the absolute FUV magnitude derived from the apparent {\sl GALEX} magnitude.}

\hbox{$^{\rm c}$$M_g$$^{\rm c}$ is the absolute SDSS $g$ magnitude corrected for the Milky Way extinction.}

\hbox{$^{\rm d}$$M_\star$ is the total stellar mass (young $+$ older population) 
in solar masses.}

\hbox{$^{\rm e}$Starburst age in Myr.}

\hbox{$^{\rm f}$$L$(H$\beta$) is the extinction-corrected H$\beta$ luminosity 
in erg s$^{-1}$.}

\hbox{$^{\rm g}$SFR is the star-formation rate in M$_\odot$ yr$^{-1}$.}

\hbox{$^{\rm h}$$\xi_{\rm ion}$ is the ionizing photon production efficiency 
in Hz erg$^{-1}$ defined as $\xi_{\rm ion}$ = $N_{\rm LyC}$/$L_\nu$, where 
$N_{\rm LyC}$ is the Lyman continuum}

\hbox{  photon production rate derived from the extinction-corrected
H$\beta$ luminosity and $L_\nu$ is the monochromatic luminosity}

\hbox{  at $\lambda$ = 1500\AA\ derived from the intrinsic 
rest-frame SED.}

\hbox{$^{\rm i}$$\alpha$ is the exponential disc scale length in kpc.}

\hbox{$^{\rm j}$$r_{50}$ is the galaxy radius in kpc where the NUV intensity is equal to half of the maximal intensity.}

\hbox{$^{\rm k}$$\Sigma_1$ is the star-formation rate surface density in M$_\odot$yr$^{-1}$kpc$^{-2}$, assuming the galaxy radius to be equal to $\alpha$.}

\hbox{$^{\rm l}$$\Sigma_2$ is the star-formation rate surface density in M$_\odot$yr$^{-1}$kpc$^{-2}$, assuming the galaxy radius to be equal to $r_{50}$.}

  \end{table*}

  \begin{table}
  \caption{{\sl HST}/COS observations \label{tab4}}
  \begin{tabular}{lccc} \hline
\multicolumn{1}{c}{Name}&\multicolumn{1}{c}{Date}&\multicolumn{2}{c}{Exposure time (s)} \\ 
    &    &MIRRORA&G130M \\ \hline
J0007$+$0226&2018-10-30&2$\times$1398     & 5616\\
J0159$+$0751&2018-01-26&2$\times$500      & 4245\\
J0820$+$5431&2019-02-25&2$\times$1494     & 6001\\
J0926$+$4504&2019-02-22&2$\times$1470     & 5905\\
J1032$+$4919&2018-11-18&2$\times$440      & 4645\\
J1205$+$4551&2019-01-03&2$\times$610      & 7470\\
J1242$+$4851&2018-12-22&2$\times$0      & 6890\\
J1355$+$4651&2019-02-22&2$\times$670      & 7350\\
\hline
\end{tabular}
  \end{table}

The emission-line luminosities and stellar masses of our galaxies were obtained 
adopting a luminosity distance \citep[NED,][]{W06} with the cosmological 
parameters $H_0$=67.1 km s$^{-1}$Mpc$^{-1}$, $\Omega_\Lambda$=0.682, 
$\Omega_m$=0.318 \citep{P14}. Using these parameters and apparent magnitudes,
we also derived absolute magnitudes $M_g$ and $M_{\rm FUV}$ in the SDSS $g$ and
{\sl GALEX} FUV bands (Table \ref{tab3}), assuming negligible bandpass 
corrections and neglecting dust extinction.

The H$\beta$ luminosity $L$(H$\beta$) and corresponding star-formation rates 
SFR shown in Table \ref{tab3} were obtained from the extinction-corrected 
H$\beta$ flux densities using the relation by \citet{K98} and adopting
the value of 2.75 for the H$\alpha$/H$\beta$ flux ratio. 
They are one order of magnitude lower than the respective values for LyC 
leakers. Specific star formation rates
sSFR = SFR/$M_\star$ of $\sim$ 100 Gyr$^{-1}$ are similar to sSFRs for LyC 
leakers \citep{I16,I16b,I18,I18b} and are among the highest known for 
GPs and luminous compact galaxies \citep*{Ca09,I11}, and dwarf SFGs
at $z < 1$ \citep{I16c}. On the other hand, they are several orders of
magnitude higher than the sSFRs of $\sim$ 0.01 - 1 Gyr$^{-1}$ of the SDSS main
sequence galaxies.

The extinction-corrected SDSS spectra of our galaxies were used to fit 
the spectral energy 
distribution (SED) of both the stellar and nebular components and to derive 
their stellar masses, starburst ages and the modelled intrinsic absolute 
magnitudes $M_{\rm FUV}^{\rm SED}$ in the FUV range. 
The fitting method, using a two-component model, is described in \citet{I18}.
We find that the stellar masses and FUV luminosities of our galaxies are 
$\sim$ 2 orders of magnitude lower compared to the confirmed LyC leakers 
(Table \ref{tab3}), but their extinction-corrected absolute FUV magnitudes are 
similar to those of the
faintest detected galaxies at $z$~=~6~--~8 \citep*[e.g. ][]{L17,Bo17b}.

\begin{figure*}
\hbox{
\includegraphics[angle=-90,width=0.24\linewidth]{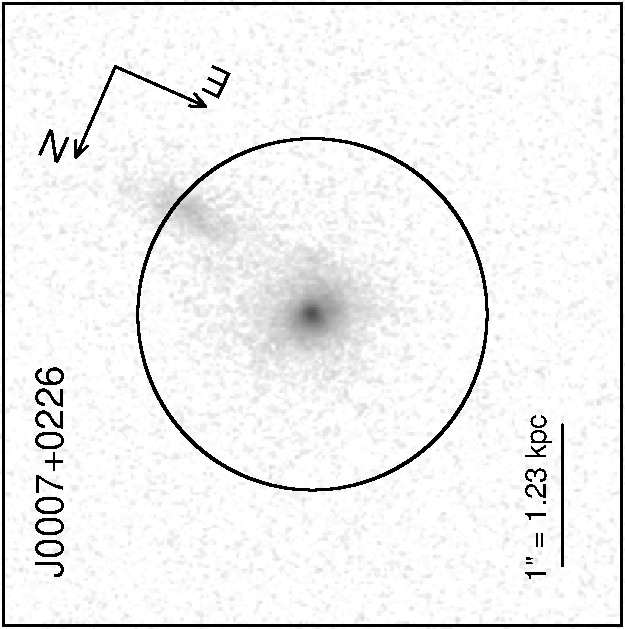}
\includegraphics[angle=-90,width=0.24\linewidth]{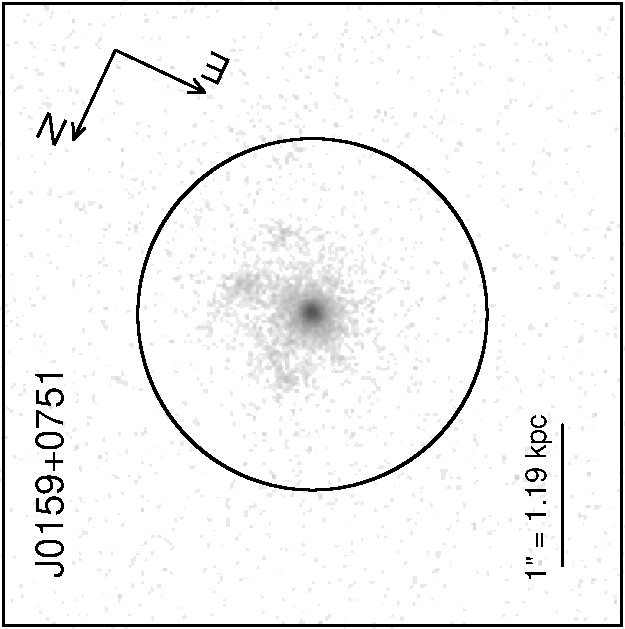}
\includegraphics[angle=-90,width=0.24\linewidth]{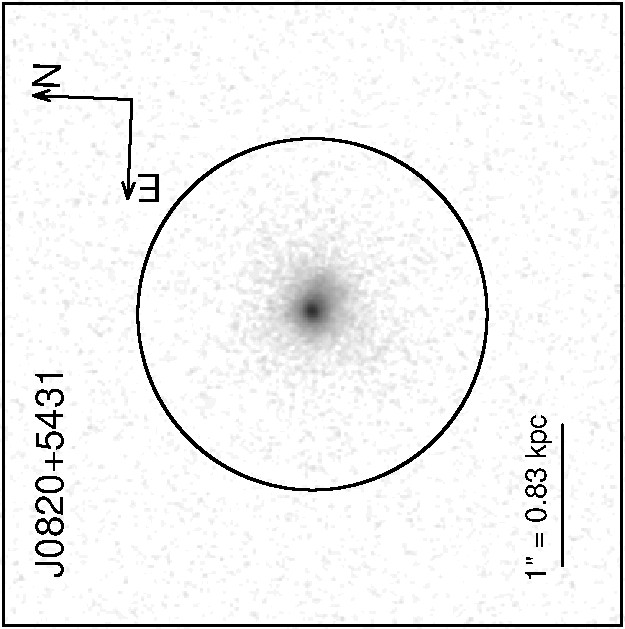}
\includegraphics[angle=-90,width=0.24\linewidth]{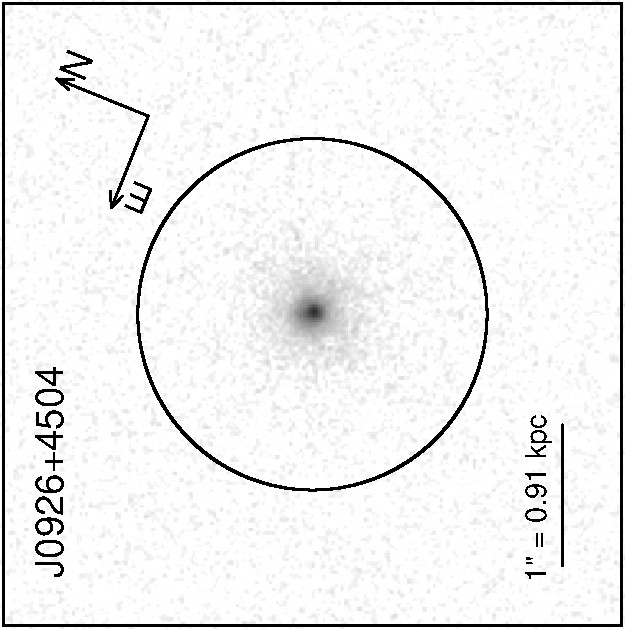}
}
\hbox{
\includegraphics[angle=-90,width=0.24\linewidth]{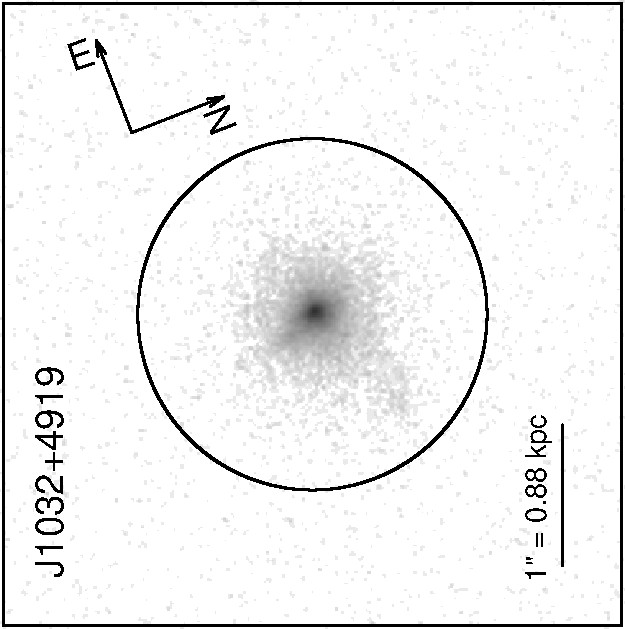}
\includegraphics[angle=-90,width=0.24\linewidth]{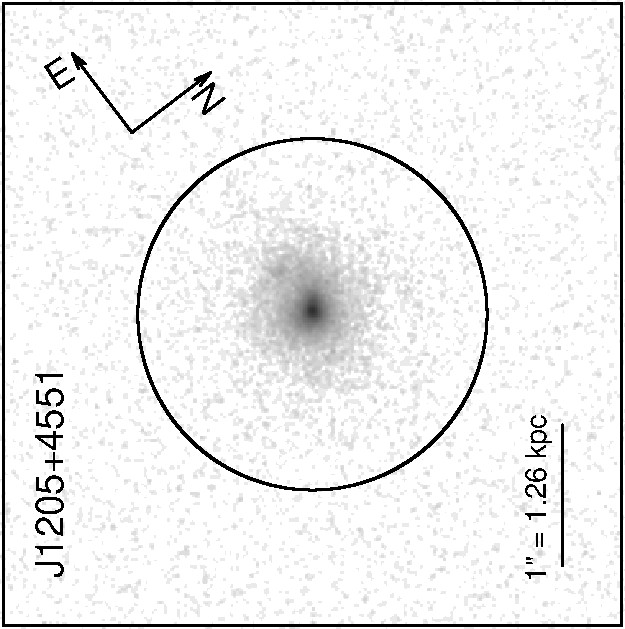}
\includegraphics[angle=-90,width=0.24\linewidth]{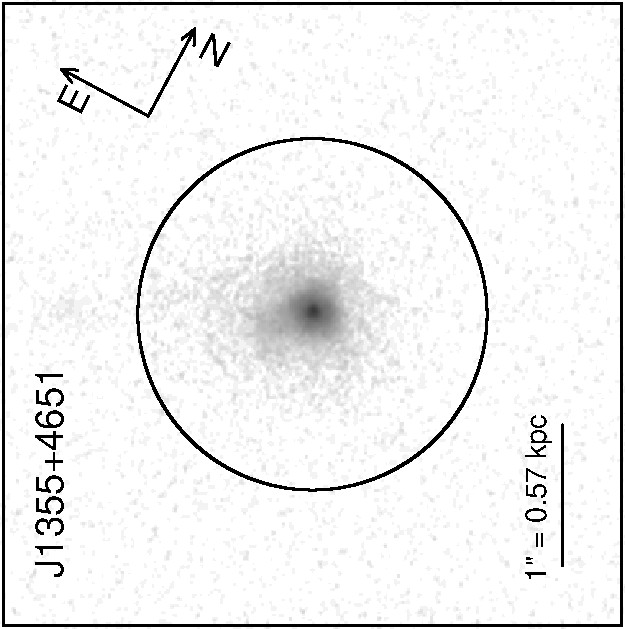}
}
\caption{The {\sl HST} NUV acquisition images of selected galaxies in 
log surface brightness scale. The image for J1242$+$4851 is not shown
(see text). The COS spectroscopic aperture with a diameter of 
2.5 arcsec is shown in all panels by a circle.
The linear scale in each panel is derived adopting the angular size distance
(Table~\ref{tab1}).
\label{fig4}}
\end{figure*}

\begin{figure*}
\hbox{
\includegraphics[angle=-90,width=0.33\linewidth]{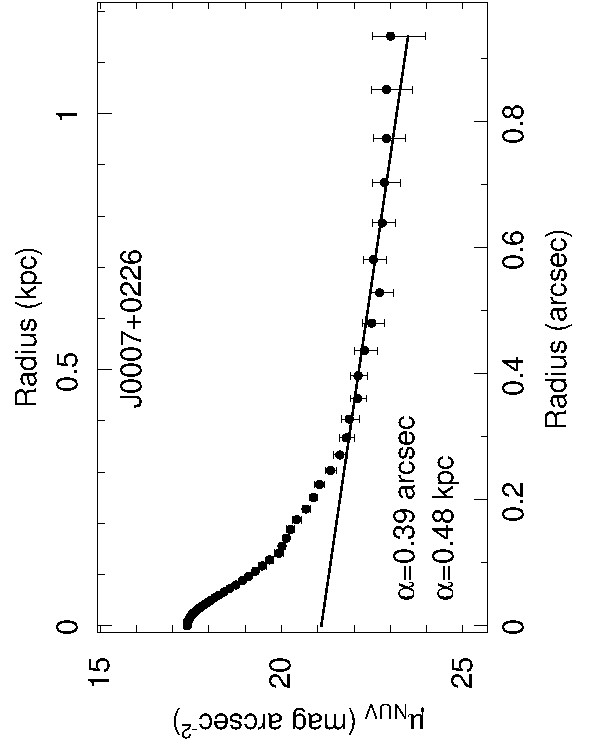}
\includegraphics[angle=-90,width=0.33\linewidth]{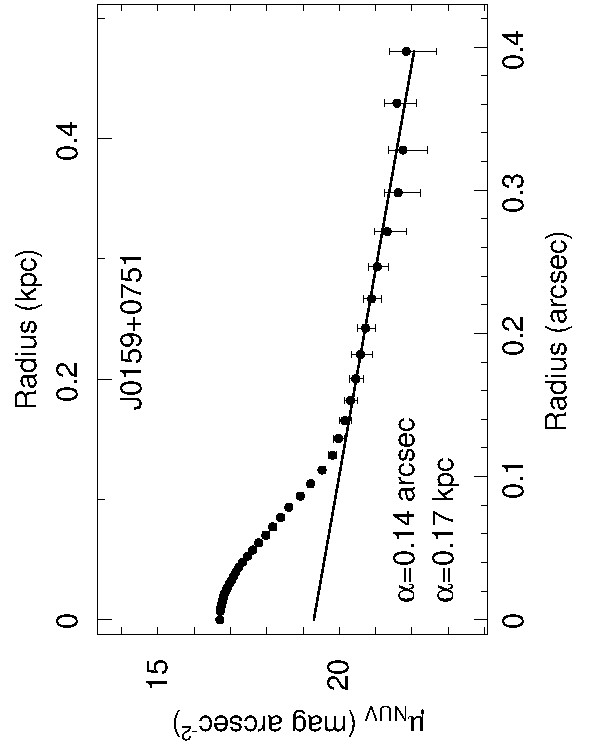}
\includegraphics[angle=-90,width=0.33\linewidth]{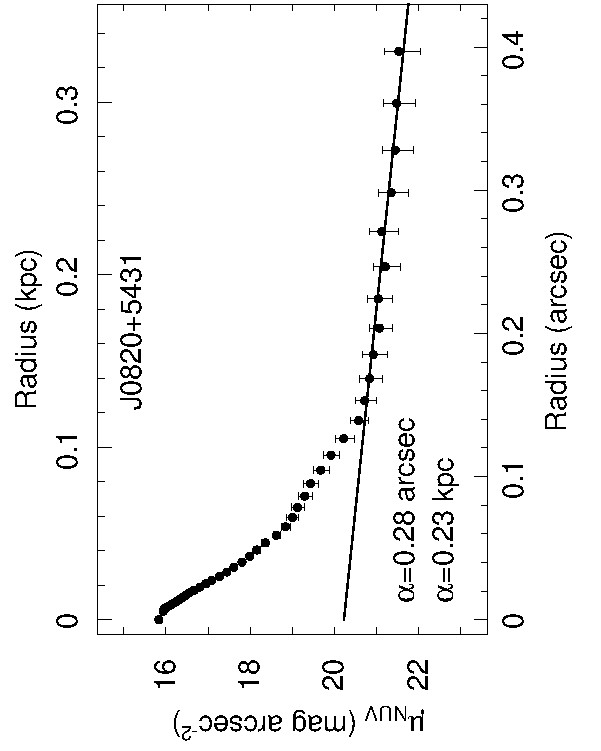}
}
\hbox{
\includegraphics[angle=-90,width=0.33\linewidth]{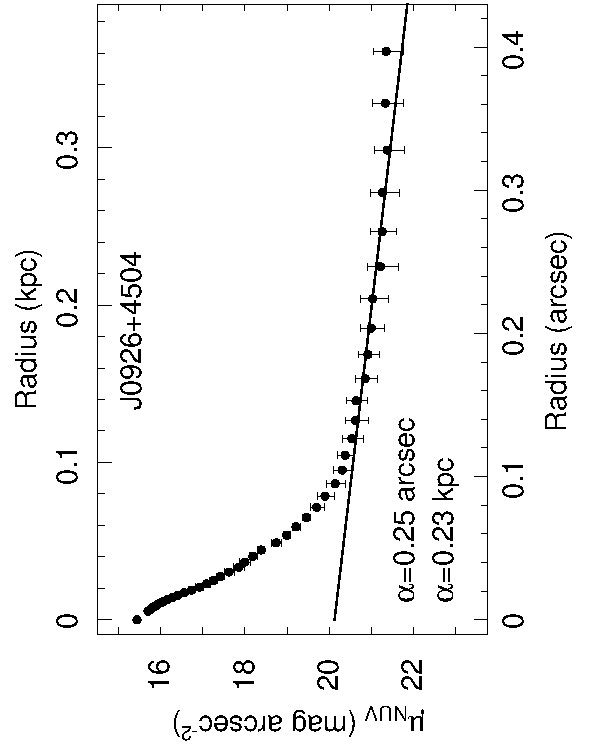}
\includegraphics[angle=-90,width=0.33\linewidth]{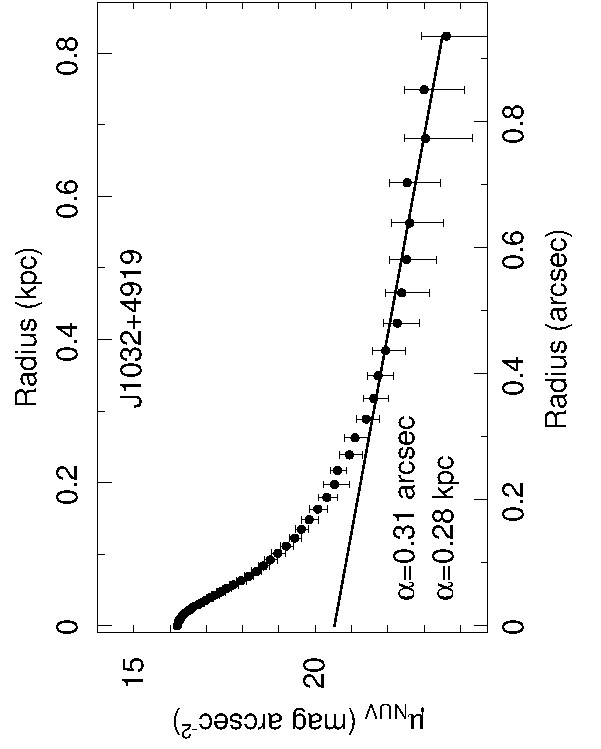}
\includegraphics[angle=-90,width=0.33\linewidth]{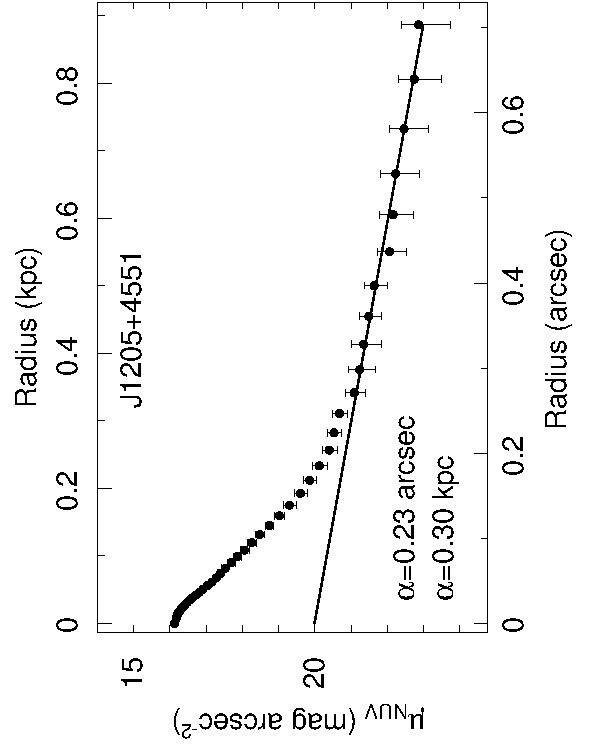}
}
\hbox{
\includegraphics[angle=-90,width=0.33\linewidth]{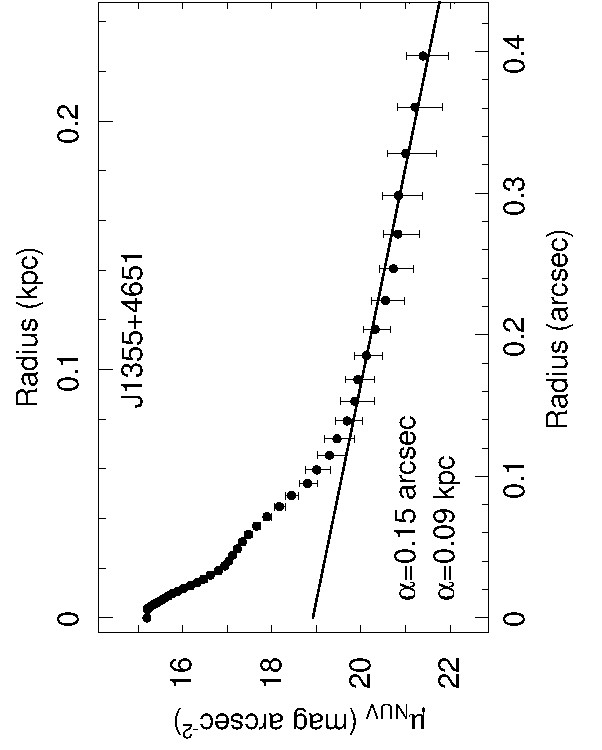}
}
\caption{NUV surface brightness profiles of galaxies with the highest
O$_{32}$ ratio. The profile for
J1242$+$4851 is not shown. The linear fits to the surface brightness profiles of
the outer galaxy parts are shown by solid lines.
\label{fig5}}
\end{figure*}

\section{{\sl HST}/COS observations and data 
reduction}\label{sec:obs}

{\sl HST}/COS spectroscopy of the eight galaxies was obtained
in program GO~15136 (PI: Y.\ I.\ Izotov) during the period January 2018 --
February 2019. The observational details
are presented in Table \ref{tab4}. The galaxies were acquired by COS near
ultraviolet (NUV) imaging. Since our targets are compact but faint, as
based on shallow {\sl GALEX} imaging, a considerable fraction of one orbit was spent per object for deep NUV imaging and reliable acquisition.
The NUV-brightest region of each target was centered in the
2.5\,arcsec diameter spectroscopic aperture (Fig. \ref{fig4}).
We note that acquisition observations failed for J1242$+$4851.
We do not know the reason for that. The acquisition image has a flag 
raised in the header: EXPFLAG=`INVALID CHECK IMAGE'. Possibly, something 
went wrong during one of the two acquisition exposures. 
Therefore, the 
imaging data for J1242$+$4851 are not present in this paper. However, the 
spectroscopic observations were not affected.
For the other galaxies, some structure with an extended
low-surface-brightness (LSB) component is seen. However, their sizes are smaller
than the central unvignetted $0.8$\,arcsec diameter region of
the spectroscopic aperture. Hence, the galaxy quantities 
derived from the COS spectra here do not require corrections for vignetting.
In fact, most of the NUV continuum in these galaxies is 
concentrated in much
smaller central regions, with full widths at half maximum $\la$ 0.1 arcsec.

We obtained spectra of the Ly$\alpha$ emission lines  
with the COS G130M grating positioned at the central wavelength 1222\,\AA\, 
and a small wavelength coverage of $\sim$ 300\,\AA\ at COS Detector 
Lifetime Position 4, yielding a spectral resolving power 
$\lambda/\Delta\lambda\simeq 14,000$ in the wavelength range of interest. All 
four focal-plane offset positions were employed to correct grid wire shadows 
and detector blemishes.

The individual exposures were reduced with the CALCOS pipeline v3.3.4, followed 
by accurate background subtraction and co-addition with custom software 
\citep{W16}. Apart from adjustments of the rectangular extraction aperture due 
to the wide spatial profile of the G130M 1222\,\AA\ setup (57 pixels) and 
adjustments in the detector pulse-height thresholds to reflect the current 
detector state, the reduction procedures were analogous to \citet{I16}. We 
verified in the two-dimensional spectra that the extraction aperture contains 
all visible Ly$\alpha$ flux and preserves the spectrophotometry of the 
continuum. The estimated background uncertainties of a few per cent do not 
affect our analysis. The homogeneous reduction enables a fair comparison to 
our previous results \citep{I16,I16b,I18,I18b}.

\begin{figure*}
\hbox{
\includegraphics[angle=-90,width=0.33\linewidth]{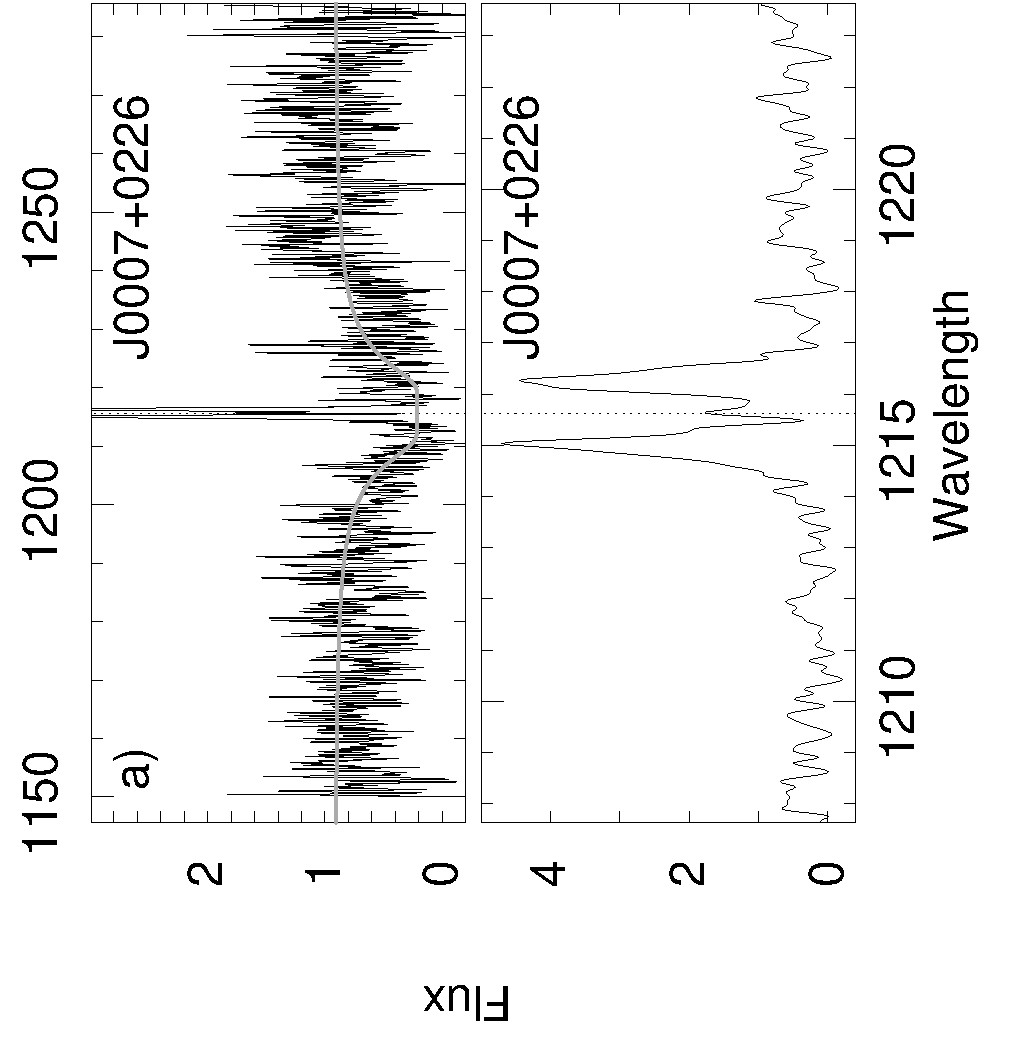}
\includegraphics[angle=-90,width=0.33\linewidth]{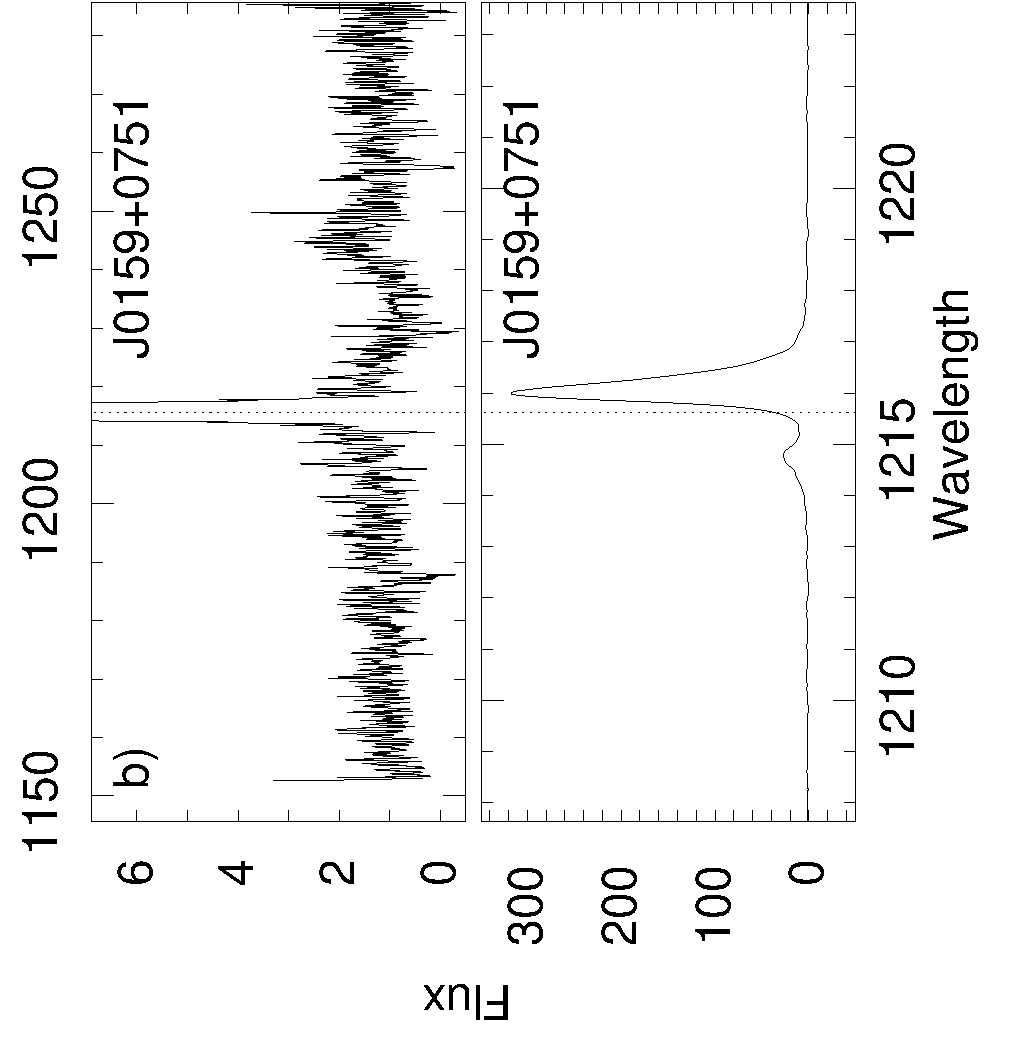}
\includegraphics[angle=-90,width=0.33\linewidth]{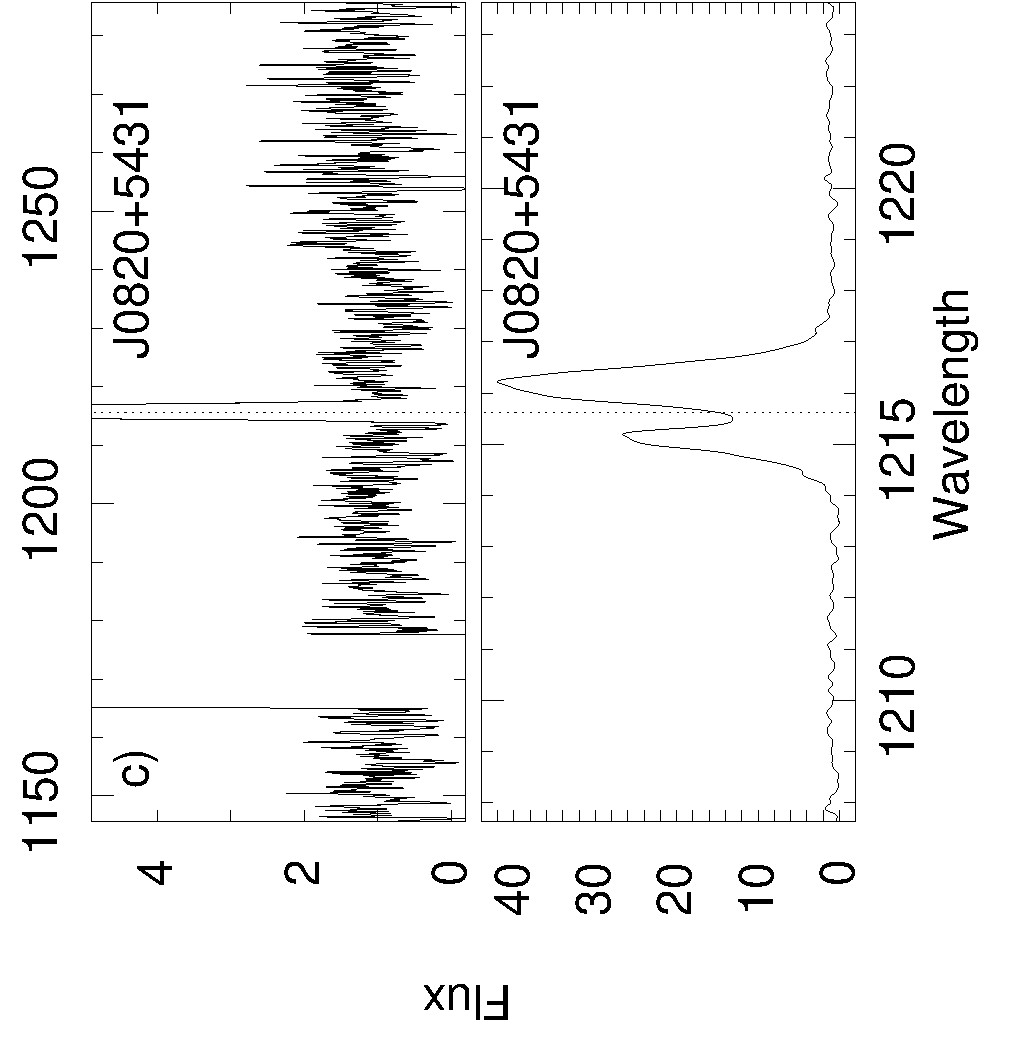}
}
\hbox{
\includegraphics[angle=-90,width=0.33\linewidth]{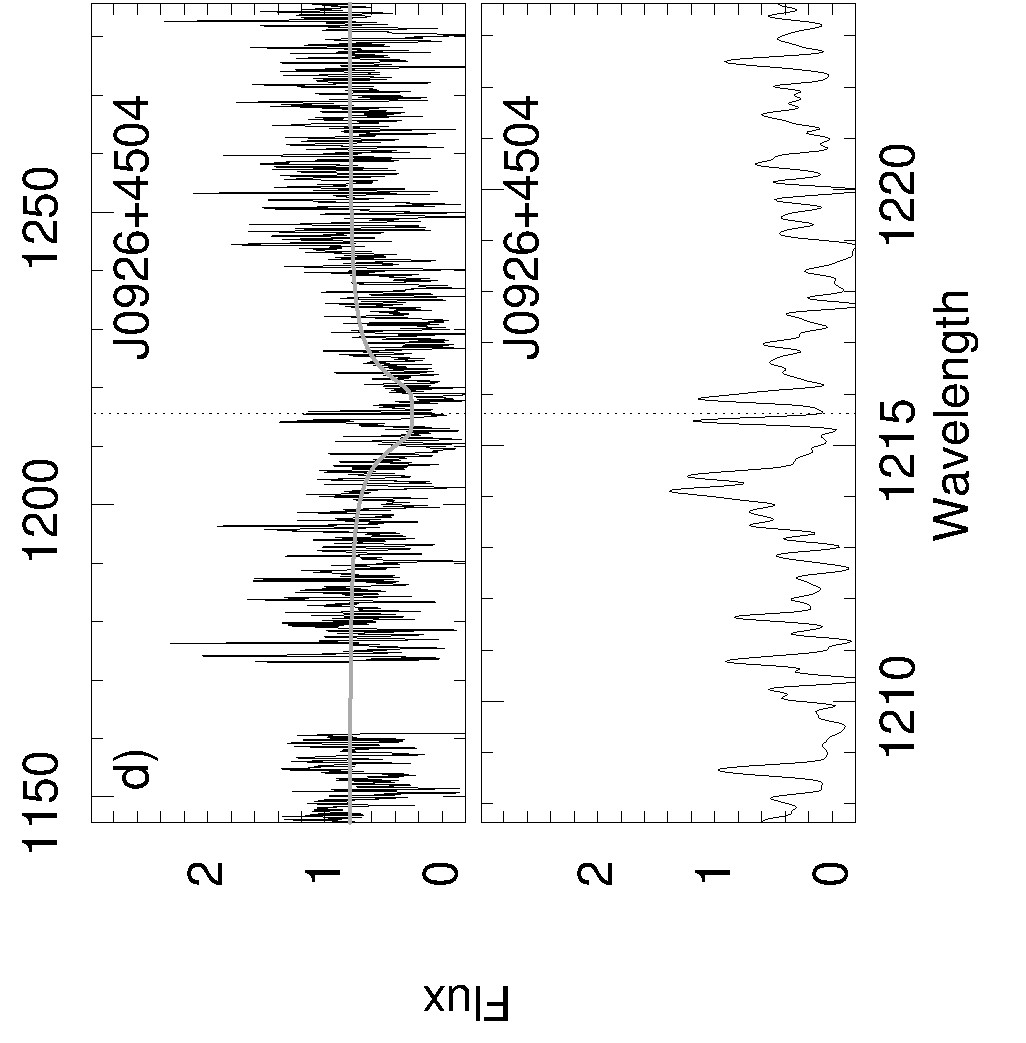}
\includegraphics[angle=-90,width=0.33\linewidth]{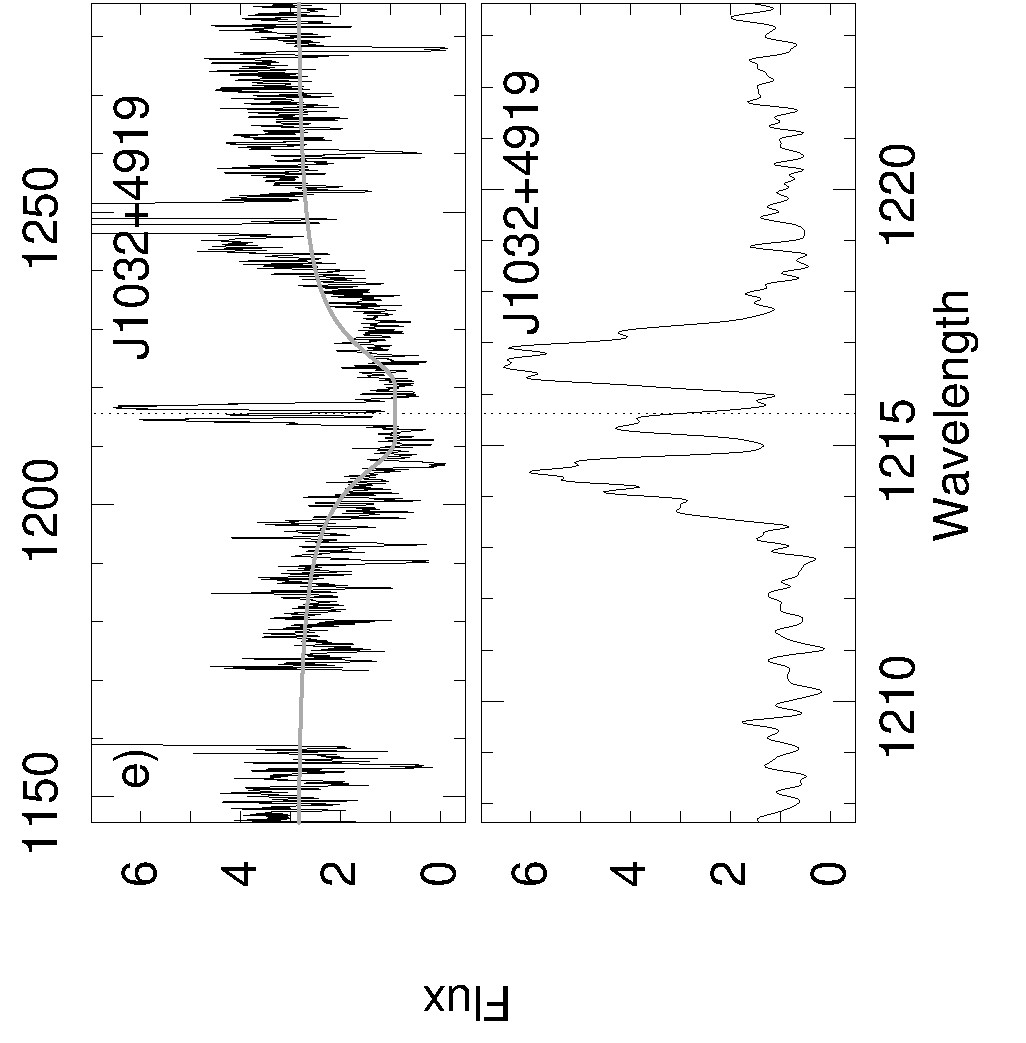}
\includegraphics[angle=-90,width=0.33\linewidth]{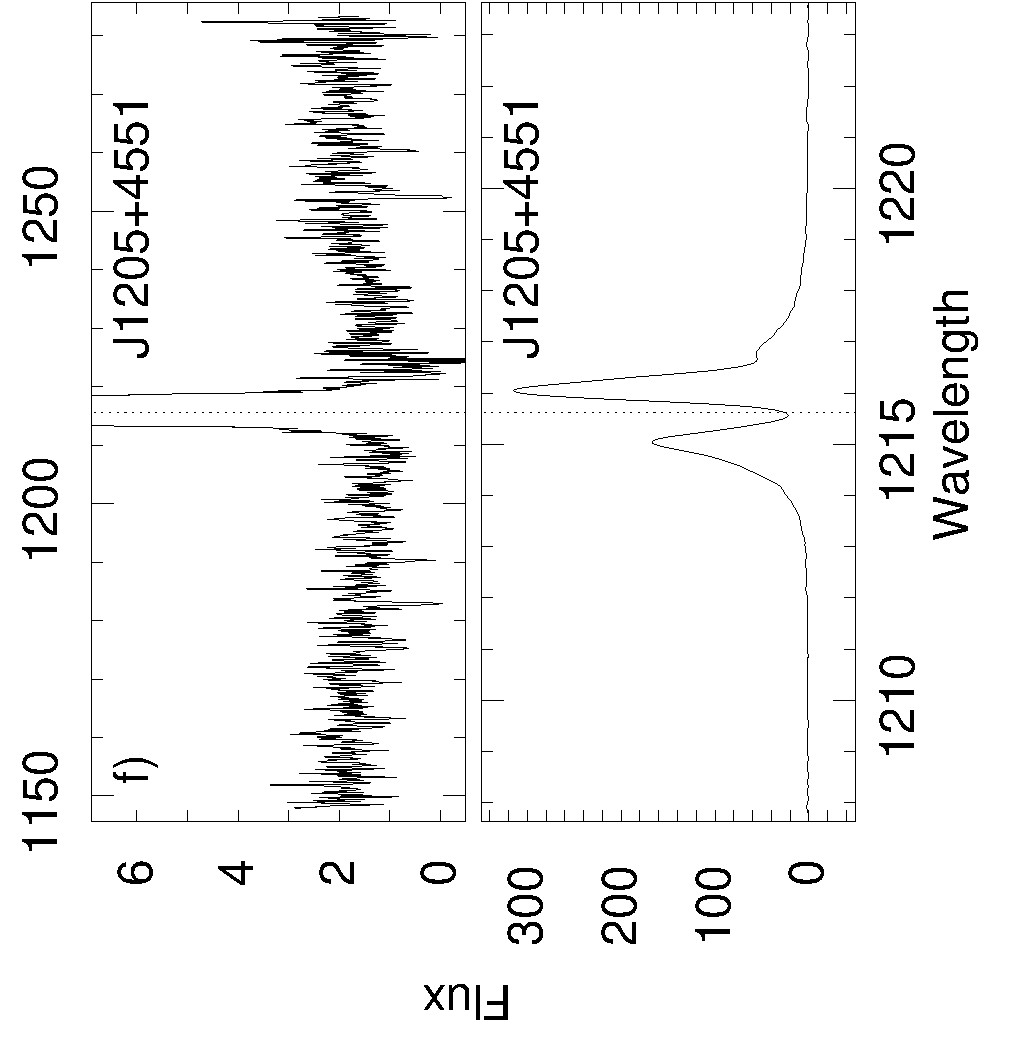}
}
\hbox{
\includegraphics[angle=-90,width=0.33\linewidth]{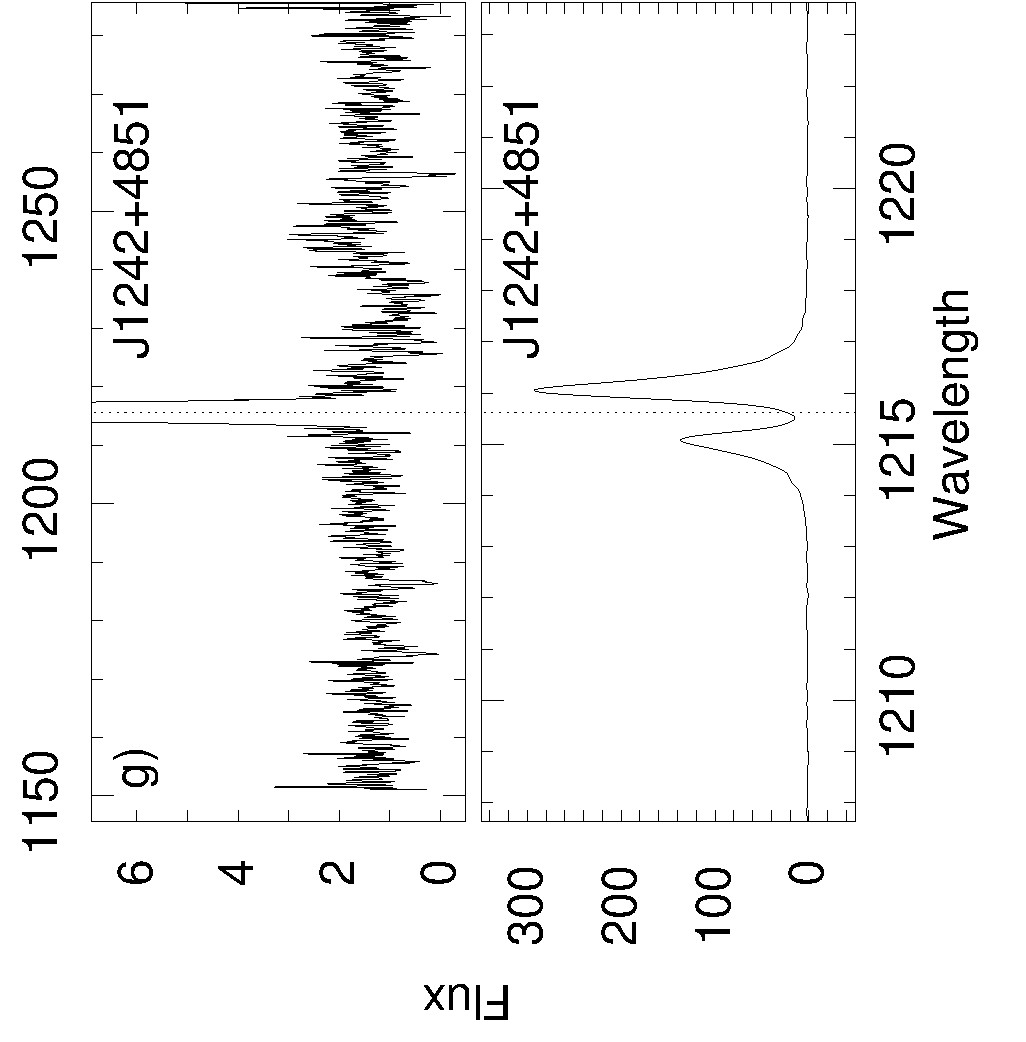}
\includegraphics[angle=-90,width=0.33\linewidth]{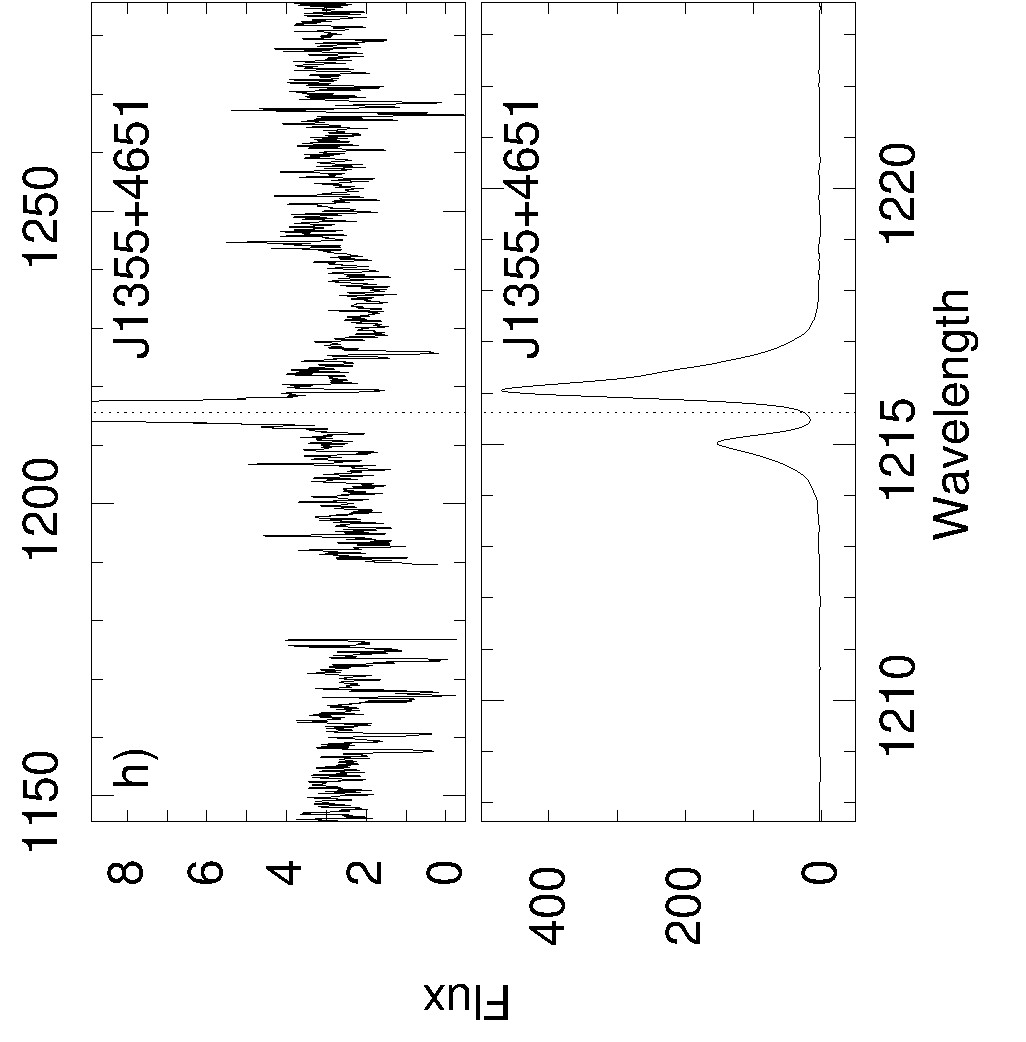}
}
\caption{Ly$\alpha$ profiles. Vertical dotted lines indicate the centres of 
profiles. Upper panels for each object, with the rest-frame wavelength scale on 
top, are adjusted so that the continuum can be seen  
in a wide range of wavelengths. Lower panels, with the rest-frame wavelength scale at bottom, are adjusted to see Ly$\alpha$ emission-line profile.
The geocoronal O~{\sc i} $\lambda$1301,1304\AA\ emission in spectra was removed 
by considering only orbital night data in the affected wavelength range. This 
wavelength range has a shorter exposure time and a lower S/N. In particular,
the depression and higher noise in the upper panel of f) for J1205$+$4551, at 
$\sim$ $\lambda$1225\AA, is caused by replacement of the total spectrum by 
the orbital night spectrum in the affected region.
O~{\sc i} $\lambda$1301,1304\AA\ emission could not be removed for J1032$+$4919.
Flux densities are in 10$^{-16}$ erg s$^{-1}$ cm$^{-2}$\AA$^{-1}$ 
and wavelengths are in \AA. \label{fig6}}
\end{figure*}

  \begin{table*}
  \caption{Parameters for the Ly$\alpha$ and H$\beta$ emission lines \label{tab5}}
  \begin{tabular}{rcrcrcrrc} \hline
Name&$A$(Ly$\alpha$)$_{\rm MW}$$^{\rm a}$&\multicolumn{1}{c}{$I$(Ly$\alpha$)$^{\rm b}$}&log $L$(Ly$\alpha$)$^{\rm c}$&\multicolumn{1}{c}{EW(Ly$\alpha$)$^{\rm d}$}&\multicolumn{1}{c}{$V_{\rm sep}$$^{\rm e}$}&\multicolumn{1}{c}{$I$(H$\beta$)$^{\rm f}$}
&$f_{\rm esc}$(Ly$\alpha$)$^{\rm g}$&$f_{\rm esc}$(LyC)$^{\rm h}$\\ 
\hline
J0007$+$0226&0.161&   6.5$\pm$3.2&39.81&    7.4$\pm$3.9& 310.3$\pm$22.5&52.3$\pm$\,~7.0&0.53$\pm$0.27& ...\\ 
J0159$+$0751&0.767& 452.1$\pm$6.7&41.61&  160.2$\pm$2.4& 311.5$\pm$\,~4.0&99.3$\pm$13.0&19.54$\pm$2.57& \,\,\,9.08$\pm$3.00\\ 
J0820$+$5431&0.438&  57.5$\pm$2.5&40.30&   47.8$\pm$2.2& 255.7$\pm$\,~3.7&20.2$\pm$\,~2.7&12.22$\pm$1.72&17.84$\pm$3.01\\ 
J0926$+$4504&0.173&   2.0$\pm$1.9&38.93&    2.2$\pm$2.1&      ...       &34.4$\pm$\,~4.5&0.25$\pm$0.24& ...\\ 
J1032$+$4919&0.070&  14.9$\pm$2.0&39.84&  5.2$\pm$0.8& 247.7$\pm$\,~9.9&109.6$\pm$14.2&0.59$\pm$0.11& ...\\ 
            &     &              &     &             & 281.3$\pm$\,~9.9&             &    &    \\ 
J1205$+$4551&0.138& 435.1$\pm$4.9&41.66&189.5$\pm$3.2& 248.8$\pm$\,~3.2&115.0$\pm$15.0&16.24$\pm$2.12&19.48$\pm$3.01\\ 
J1242$+$4851&0.106& 293.9$\pm$3.9&41.44&191.3$\pm$2.5& 239.4$\pm$\,~4.9&40.1$\pm$\,~5.5&31.46$\pm$4.34&22.00$\pm$3.03\\ 
J1355$+$4651&0.156& 461.9$\pm$2.9&40.93&175.7$\pm$1.2& 259.9$\pm$\,~4.5&51.2$\pm$\,~6.7&38.72$\pm$5.07&16.92$\pm$3.01\\ 
\hline
  \end{tabular}

\hbox{$^{\rm a}$$A$(Ly$\alpha$)$_{\rm MW}$ is the Milky Way extinction at the observed wavelength of the Ly$\alpha$
emission line in mags, adopting \citet{C89}}

\hbox{\, reddening law with $R(V)$=3.1.}

\hbox{$^{\rm b}$$I$(Ly$\alpha$) is the Ly$\alpha$ flux density in 10$^{-16}$ erg s$^{-1}$ cm$^{-2}$ measured in
the COS spectrum and corrected for the Milky Way extinction.}

\hbox{$^{\rm c}$$L$(Ly$\alpha$) is the Ly$\alpha$ luminosity in erg s$^{-1}$ corrected for the
Milky Way extinction.}

\hbox{$^{\rm d}$EW(Ly$\alpha$) is the rest-frame equivalent width in \AA\ of the Ly$\alpha$ emission line. Rest-frame equivalent widths of 
Ly$\alpha$ absorption lines in} 

\hbox{\, J0007$+$0226, J0926$+$4504, and J1032$+$4919 are $-$15.7$\pm$5.1~\AA, $-$20.6$\pm$7.3~\AA, and $-$21.4$\pm$3.1~\AA, respectively.}


\hbox{$^{\rm e}$$V_{\rm sep}$ is the Ly$\alpha$ velocity peak separation in km s$^{-1}$.}


\hbox{$^{\rm f}$$I$(H$\beta$) is the extinction-corrected H$\beta$ flux density in 
10$^{-16}$ erg s$^{-1}$ cm$^{-2}$ measured in the SDSS spectrum.}

\hbox{$^{\rm g}$$f_{\rm esc}$(Ly$\alpha$) is the ratio in percentage of 
$I$(Ly$\alpha$)/$I$(H$\beta$) to its case B value of 23.3.}

\hbox{$^{\rm h}$$f_{\rm esc}$(LyC) is the indirectly derived LyC escape fraction in
percentage, using the value of $V_{\rm sep}$ and equation 2 in \citet{I18b}.}

  \end{table*}

\section{Surface brightness distribution in the NUV range}\label{sec:sbp}

Using the COS NUV acquisition images and the routine {\it ellipse} in 
{\sc iraf}\footnote{{\sc iraf} is distributed by 
the National Optical Astronomy Observatories, which are operated by the 
Association of Universities for Research in Astronomy, Inc., under cooperative 
agreement with the National Science Foundation.}/{\sc stsdas}\footnote{{\sc stsdas} is a product of 
the Space Telescope Science Institute, which is operated by AURA for NASA.} 
we obtain the surface brightness (SB) profiles of our galaxies in the UV 
continuum. The {\sl GALEX} NUV magnitudes were used to express the profiles in 
the units mag arcsec$^{-2}$. For this, we have scaled the total NUV flux of
the galaxy measured in the acquisition image to the flux corresponding to
the {\sl GALEX} NUV magnitude. The shape of the SB profiles (Fig.~\ref{fig5}) 
in our galaxies is similar to those of confirmed LyC leakers 
\citep{I16,I16b,I18,I18b}
with a sharp increase in the central part because of the presence
of the bright star-forming region(s) and 
a linear decrease (in magnitudes) in the outer part, 
reminiscent of a disc structure.

However, the scale lengths $\alpha$ of our galaxies in the range 
$\sim$ 0.09 -- 0.48 kpc (Fig.~\ref{fig5}) are considerably lower than 
$\alpha$ = 0.6 -- 1.8 kpc in LyC leakers \citep{I16,I16b,I18,I18b},
presumably because of their considerably lower stellar masses. 
The scale lengths of our galaxies are among the lowest found for local
blue compact dwarf galaxies \citep{P02}. On the other hand, the 
corresponding surface densities of SFR in the studied galaxies 
$\Sigma$ = SFR/($\pi \alpha^2$) are comparable to those of LyC leakers. 
Because of the compactness of the bright star-forming regions, the 
half-light radii $r_{50}$ of our galaxies in the NUV are 
considerably smaller than $\alpha$ (see Table \ref{tab3}).
Adopting $r_{50}$ as a measure of the size of these galaxies (Table \ref{tab3}),
the corresponding $\Sigma$s are
typically two orders of magnitude larger, and comparable to those found for 
low-redshift LyC leakers and SFGs at high redshifts \citep{CL16,PA18,Bo17}.

\section{Ly$\alpha$ emission}\label{sec:lya}

The LyC emission can not be directly observed in our galaxies because of their 
low redshifts. Therefore, we wish to use the Ly$\alpha$ profiles in their 
spectra to derive information 
about possible leaking LyC radiation for galaxies with much
lower stellar masses than those of confirmed LyC leakers. 
\citet{D16} and \citet{V17} have proposed that the presence of a 
double-peaked Ly$\alpha$ profile, with a small peak separation, would be a good 
indicator of LyC leakage. According to the models 
of \citet{V15}, the peak separation decreases with decreasing 
column density of the neutral gas. This in turn would result in a higher escape
fraction of the LyC radiation. Additionally, the role of the blue Ly$\alpha$ 
peak in the LyC escape has been discussed by e.g. \citet{H15} and \citet{O18}.
They found that the escape fraction of the Ly$\alpha$ radiation increases
when the blue peak velocity decreases.

The Ly$\alpha$ profiles in the medium-resolution spectra of the 8 
observed galaxies are shown in 
Fig. \ref{fig6}. In the upper panel for each object, both
the flux and the wavelength are scaled so that the continuum 
behaviour can be seen over a broader wavelength range around the Ly$\alpha$ line, while in 
the lower panel both the flux and the wavelength scales are adjusted to examine  
the profile of the Ly$\alpha$ emission line.
A strong Ly$\alpha$ $\lambda$1216\,\AA\ emission-line without any evidence for 
an absorption profile is detected in five galaxies 
(Fig. \ref{fig6}b,c,f,g,h), similarly to the low-$z$ LyC leakers 
\citep[fig. 7 in ][]{V17}. In the remaining 3 galaxies, the weak Ly$\alpha$
emission line is superposed on a broad absorption component
(Fig. \ref{fig6}a,d,e). The fraction of galaxies with a detected Ly$\alpha$
absorption line is similar to that obtained by \citet{MK19}.
However, the intensity at the bottom of the absorption
lines is above zero in these galaxies, indicating that the covering factor 
of the H~{\sc i} cloud surrounding the ionizing source is less than unity,
allowing for some Ly$\alpha$ emission and some continuum emission in this 
wavelength range to escape. The full widths at zero intensity of the 
Ly$\alpha$ emission lines in all our galaxies are FWZI = 3 -- 4\AA\ 
(Fig. \ref{fig6}). On the other hand, the widths of the absorption lines at the
bottom of their profiles are considerably broader, $\sim$ 10 -- 20\AA\ in the 
three galaxies J0007$+$0226, J0926$+$4504 and J1032$+$4919. 
Therefore, it is unlikely that the emission above zero at the bottom of the 
absorption profiles is due to the Ly$\alpha$ emission line.

Six out of eight observed galaxies show double-peaked Ly$\alpha$ emission 
profiles with the peak separations much larger than the nominal COS 
spectral resolution of $\sim$ 20 km s$^{-1}$ for the G130M grating. 
The Ly$\alpha$ emission line in J0926$+$4504 is too
weak to definitely derive its profile.
The two-peak shape is similar to that observed in known LyC 
leakers \citep{I16,I16b,I18,I18b,V17} and in GP galaxies 
\citep{JO14,H15,Y17,MK19}. 
The Ly$\alpha$ emission-line profile of J1032$+$4919 is more complex and 
consists of three peaks, similar to that of one low-$z$ LyC leaker \citep{I18b},
and two $z$ $>$ 2 LyC leakers \citep{Va18,RT17}. However, at 
variance with J1032$+$4919, all other triple-peaked spectra in the literature 
do not show clear signs of underlying Ly$\alpha$ absorption.

Some parameters of the Ly$\alpha$ emission profile
are presented in Table~\ref{tab5}. The Ly$\alpha$ 
emission-line fluxes in the spectra of galaxies with detected broad Ly$\alpha$
absorption are measured placing the continuum at the bottom of the absorption
profiles. For J1032$+$4919, two separations
(blue peak - centre peak and centre peak - red peak) are
given. It is seen that the separation between the peaks is in a narrow 
range $\sim$ 240 -- 310 km s$^{-1}$, indicating low neutral hydrogen column 
densities $N$(H~{\sc i}) and implying considerable fractions of escaping LyC 
emission. However, in the spectra of three galaxies, J0007$+$0226, J0926$+$4504 
and J1032+4919, strong Ly$\alpha$ absorption is present as well.
The flux in the central part of absorption profiles is above zero, 
although their shape indicates that they are clearly saturated. 
This appearance is in contrast to expectations for a single source surrounded 
by an uniform neutral gas cloud with high $N$(H~{\sc i}). 
We derive $N$(H~{\sc i}) using the relation
\begin{equation}
I(\lambda)=I_{\rm cont} \exp{[-\tau(\lambda)]},  \label{profile}
\end{equation}
where $I_{\rm cont}$ and $I(\lambda)$ are the flux of the continuum 
and the flux in the line at the wavelength $\lambda$, respectively,
after subtraction of the flux at the bottom of the Ly$\alpha$ absorption line.
The optical depth $\tau$($\lambda$) in the Ly$\alpha$ profile at the 
wavelength $\lambda$ is defined by the relation of \citet{B75}  
\begin{equation}
\tau(\lambda) = 4.26\times 10^{-20} \frac{N({\rm H~I})}{6.04\times 10^{-10} + (\lambda - \lambda_0)^2}, \label{Lya}
\end{equation}
where $N$(H~{\sc i}) is in cm$^{-2}$, $\lambda$ is in \AA, 
$\lambda_0$ = 1215.67~\AA, 
Using Eqs.~\ref{profile} and \ref{Lya} we find 
$N$(H~{\sc i}) = 1.8$\times$10$^{21}$ cm$^{-2}$, 
1.0$\times$10$^{21}$ cm$^{-2}$ and 3.0$\times$10$^{21}$ cm$^{-2}$
from the fit of wings of absorption Ly$\alpha$ profiles 
in the spectra of J0007$+$0226, J0926$+$4504 and J1032$+$4919, respectively
(grey lines in the upper panels of Fig. \ref{fig6}a, \ref{fig6}d, and
\ref{fig6}e). We note that only the blue damped wing was used for fitting
because the red wing is contaminated by the stellar N~{\sc v} $\lambda$1240\AA\ 
line with a P Cygni profile. Then the superposition of the Ly$\alpha$ absorption and
emission lines could be explained by the emission of two star-forming regions, where
the fainter one, with a flux in the continuum corresponding to the one 
at the bottom
of the absorption profile, is surrounded by neutral gas with low $N$(H~{\sc i}), 
and the brighter one is surrounded by a neutral optically thick cloud. 
The flux of the continuum for the brighter region is determined by the 
difference between the observed flux outside the absorption profile and the 
flux at its bottom. The presence of Ly$\alpha$ emission in galaxies with broad 
Ly$\alpha$ absorption could also be explained by low-column-density holes in a 
thick cloud of neutral hydrogen.

\begin{figure}
\includegraphics[angle=-90,width=0.99\linewidth]{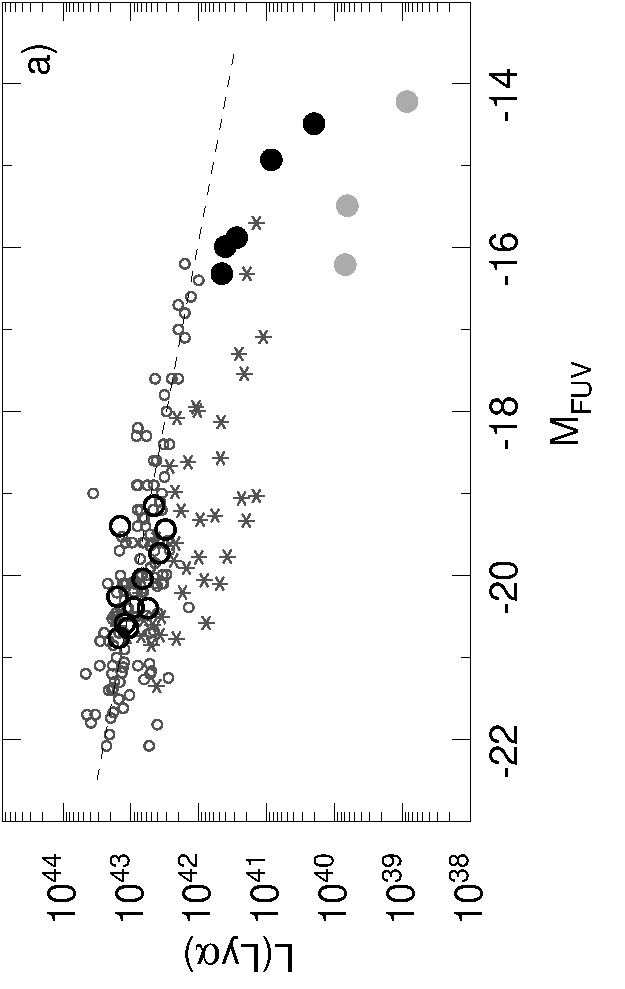}
\includegraphics[angle=-90,width=0.99\linewidth]{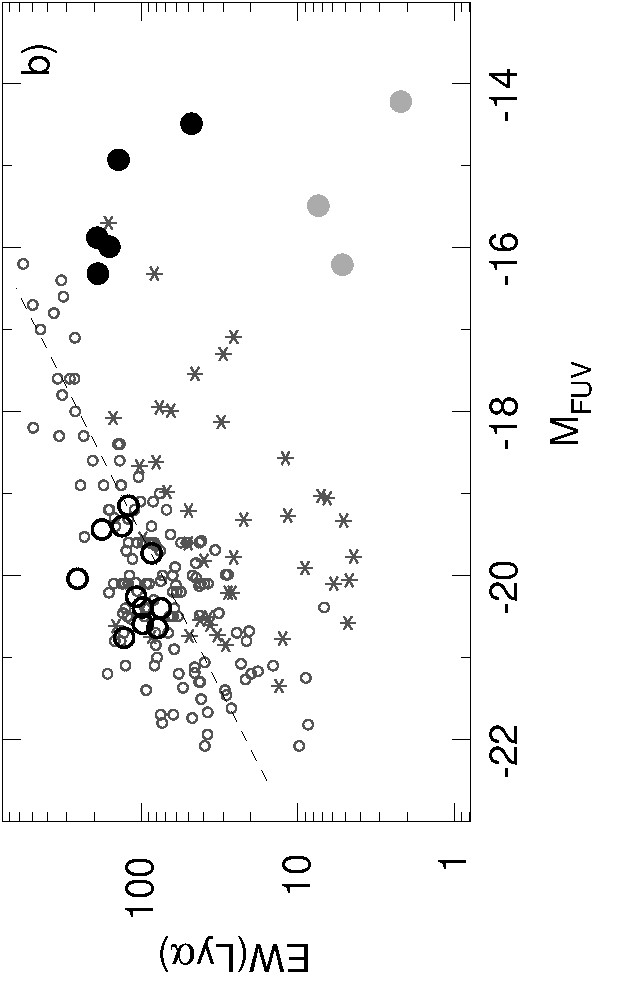}
\caption{Dependence of absolute magnitude $M_{\rm FUV}$ at 1500\AA\ on 
{\bf a)} Ly$\alpha$ luminosity $L$(Ly$\alpha$) in erg s$^{-1}$ and
{\bf b)} equivalent width EW(Ly$\alpha$) in \AA. 
Galaxies from this paper with and without broad Ly$\alpha$ absorption profiles 
are shown by grey filled circles and black filled circles, 
respectively. The confirmed LyC leakers
from \citet{I16,I16b,I18,I18b} are shown by black open circles and GPs
from {\sl HST} programs GO~12928 \citep{H15}, GO~13293 \citep{JO14},
GO~14080 \citep{J17,MK19}, and GO~14201 \citep{Y17} are shown by grey asterisks.
High-redshift galaxies from \citet{O08}, \citet{Ha17}
and \citet{J13} are represented by small grey open circles. 
Relations obtained in this paper for high-redshift galaxies 
are shown by dashed lines in both panels. \label{fig7}}
\end{figure}

\begin{figure}
\includegraphics[angle=-90,width=0.99\linewidth]{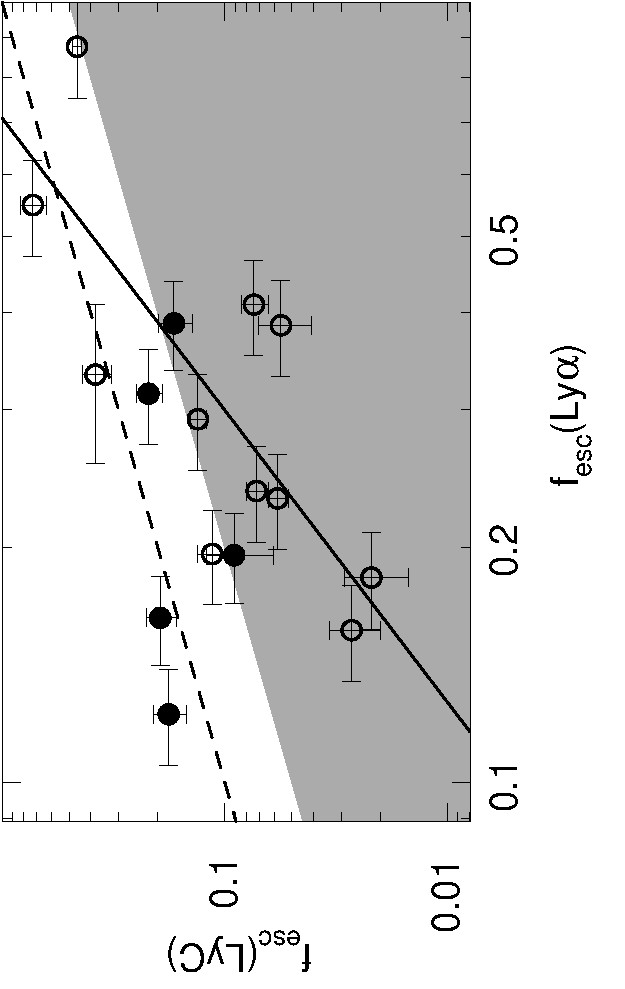}
\caption{Relation between the Ly$\alpha$ escape fraction 
$f_{\rm esc}$(Ly$\alpha$) and the directly derived LyC escape fraction 
$f_{\rm esc}$(LyC) for LyC leakers from \citet{I16,I16b,I18,I18b}
(open circles with 1 $\sigma$ error bars). The maximum likelihood fit to
these data is shown by the solid line. The dashed line shows the relation
$f_{\rm esc}$(Ly$\alpha$) = $f_{\rm esc}$(LyC). The five galaxies without 
Ly$\alpha$ absorption from this paper are represented by the filled circles. 
The LyC escape fractions for these galaxies $f_{\rm esc}$(LyC) are 
indirectly derived from the peak velocity separations of Ly$\alpha$ profiles 
(Table~\ref{tab5}),  
The shaded area shows the region occupied by the dusty clumpy models of 
\citet{D16}.
\label{fig8}}
\end{figure}

\begin{figure}
\includegraphics[angle=-90,width=0.99\linewidth]{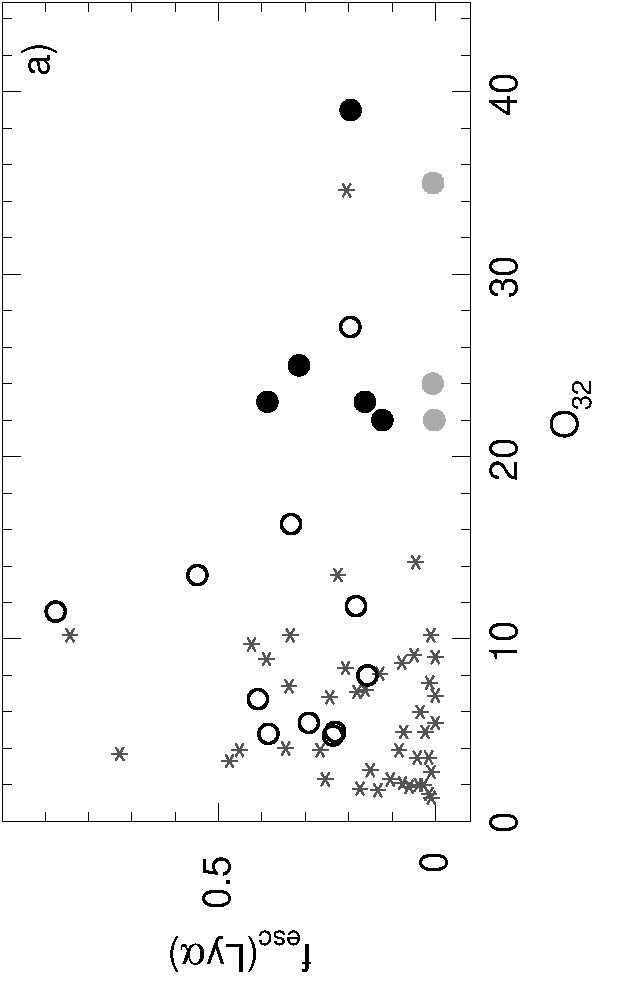}
\includegraphics[angle=-90,width=0.99\linewidth]{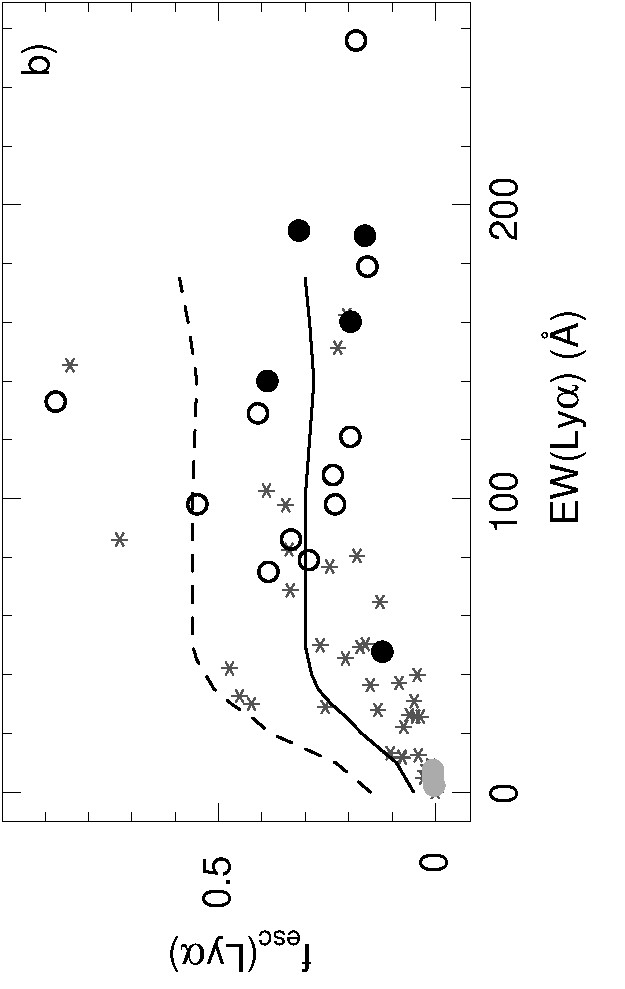}
\includegraphics[angle=-90,width=0.99\linewidth]{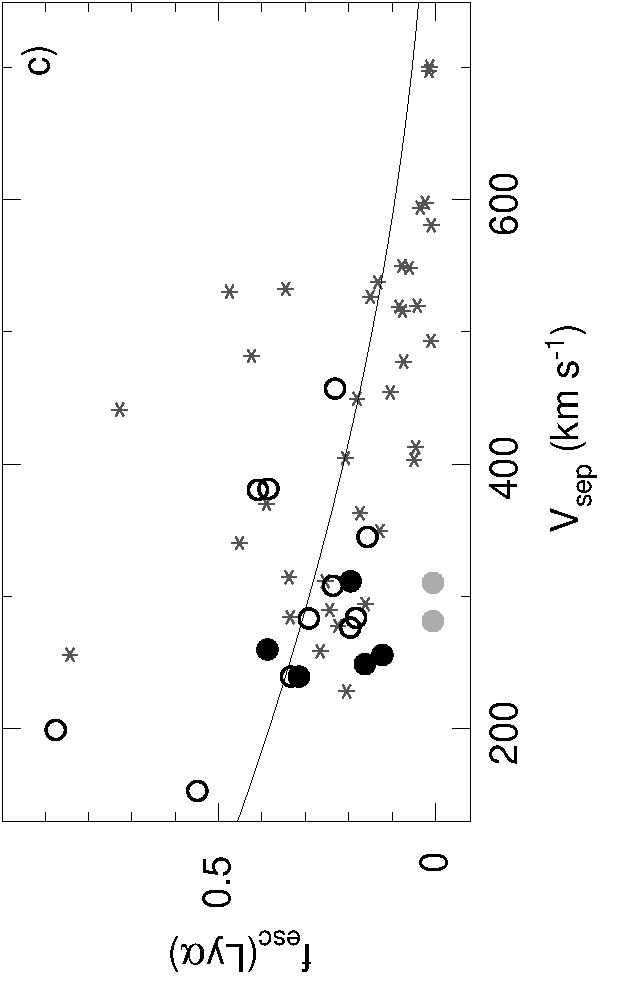}
\caption{Dependence of the Ly$\alpha$ escape fraction $f_{\rm esc}$(Ly$\alpha$) 
on {\bf a)} the [O~{\sc iii}]$\lambda$5007/[O~{\sc ii}]$\lambda$3727 flux ratio
O$_{32}$,
{\bf b)} the rest-frame Ly$\alpha$ equivalent width EW(Ly$\alpha$), and {\bf c)}
the separation between Ly$\alpha$ emission-line velocity peaks $V_{\rm sep}$.  
Symbols are the same as in Fig.~\ref{fig7}. Solid and dashed lines in {\bf b)} 
are modelled relations by \citet{SM19} for a fixed $E(B-V)$ = 0.2 and 0.1,
respectively. 
The solid line in {\bf c)} is the maximum likelihood fit to all data
(Eq. \ref{fesc1}).
\label{fig9}}
\end{figure}

\section{Discussion}\label{discussion}

\subsection{Comparison with other galaxy samples}

We compare our galaxies with those studied in other recent papers.
All galaxies from the sample in this paper are dwarf systems with SDSS $g$
absolute magnitudes $M_g$ $\sim$ --16 to --18 mag. Their absolute FUV 
magnitudes $M_{\rm FUV}$, in the range --14 to --16, compare well 
with the average $M_{\rm FUV}$ $\sim$ --15 mag of the faintest spectroscopically 
confirmed LAEs at redshifts 2.9 - 6.7, similar to those of the sources believed 
to reionize the Universe
\citep{Ma18}. Our galaxies are much more compact in the NUV continuum 
(Fig.~\ref{fig5}) and have lower metallicities (Table~\ref{tab1}) 
than most Ly$\alpha$ emitters (LAEs) and LyC leakers.
They have high ionizing photon production efficiencies $\xi_{\rm ion}$
(Table \ref{tab3}), above the canonical value in log scale of 
$\log\xi_\mathrm{ion}\sim 25.2$ in units of log Hz erg$^{-1}$
adopted in high-$z$ studies \citep[e.g. ][]{R13} and similar to the values in
LAEs at $z$ $\sim$ 3 - 7 \citep{Ha18,Na18}. These high values are a
natural consequence of a very young starburst age (Table \ref{tab3}). 
High $\xi_{\rm ion}$ in our galaxies are also similar to those in confirmed 
low-$z$ LyC leakers \citep[e.g. ][]{S16}. If applicable to all high-$z$ 
galaxies, such $\xi_{\rm ion}$ values would suffice to reionize the Universe with 
a constant escape fraction $f_{\rm esc}$(LyC) $\sim$ 10 per cent 
\citep[e.g. ][]{O09,R13,D15,Robertson15,K16}. However, we note that the 
$\xi_{\rm ion}$ values in Table \ref{tab3} should be considered only as lower
limits, as including possible escaping LyC radiation, which is not taken into 
account, would somewhat increase these values.

The high-$z$ LAEs show a tight relation between the Ly$\alpha$ luminosity
$L$(Ly$\alpha$) and the absolute magnitude $M_{\rm FUV}$ at $\lambda$ = 1500\AA\
(small open grey circles in Fig. \ref{fig7}a). The confirmed LyC leakers 
\citep[open black circles, ][]{I16,I16b,I18,I18b} follow the same relation.
On the other hand, GPs at redshifts $\la$ 0.3 (grey asterisks) are offset to 
lower $L$(Ly$\alpha$) by one order of magnitude. This offset can be
caused by aperture effects because GPs are on average closer to us and have
larger angular sizes. Furthermore, the EW(Ly$\alpha$)s in GPs are considerably 
lower than in low-$z$ LyC leakers (Fig. \ref{fig7}a), indicating older burst 
ages and thus lower Ly$\alpha$ luminosities.
We note that for the LyC leakers and GPs we do not use the published values 
for Ly$\alpha$ 
in \citet{I16,I16b,I18,I18b}, \citet{JO14}, \citet{H15}, \citet{J17} and 
\citet{Y17}. Instead for the sake of data homogeneity, we measured Ly$\alpha$ 
fluxes and equivalent widths using the public {\sl HST} data, correcting 
the fluxes for Milky Way extinction and converting them to luminosities. As for 
the absolute FUV magnitudes, we derived them from the apparent
{\sl GALEX} FUV magnitudes.

Our galaxies with extreme O$_{32}$ extend the range of $M_{\rm FUV}$ to 
appreciably lower FUV brightesses. Their low Ly$\alpha$ luminosities 
($L$(Ly$\alpha$)$\la$10$^{42}$ erg s$^{-1}$)
make the $L$(Ly$\alpha$)-$M_{\rm FUV}$ relation bend downwards at low $M_{\rm FUV}$
(Fig. \ref{fig7}a)). We note that LAEs with low Ly$\alpha$ luminosities, 
similar to those of our galaxies,
are also detected in the high-$z$ Universe \citep[e.g. ][]{L19}. 
A notable feature of the four galaxies with the highest EW(Ly$\alpha$) and 
highest $L$(Ly$\alpha$) in our sample is that their Ly$\alpha$
luminosities are lower by a factor $>$ 2 than the value derived from
the extrapolation of the relation for high-$z$ 
galaxies to fainter absolute FUV magnitudes (dashed line).
While they are consistent with the extrapolation of the trend defined by the 
GPs to fainter absolute UV magnitudes, the 
Ly$\alpha$ equivalent widths of the galaxies with extreme O$_{32}$
are similar to those of the low-$z$ LyC leakers \citep{I16,I16b,I18,I18b} 
and are much higher than those of GPs. 

It was shown e.g. by \citet{Wi16,Wi18} that high-$z$ galaxies at $z$ $>$ 3
are surrounded by extended diffuse Ly$\alpha$-emitting haloes tracing neutral 
gas up to several
tens of kpc and contributing 40 -- 90 per cent of the total Ly$\alpha$ flux. 
These extended haloes are seen even in low-mass SFGs with stellar masses down to
10$^{8}$ -- 10$^{9}$ M$_\odot$ \citep*{E18,E19}.
\citet{Y17b} have shown that Ly$\alpha$ emission in Ly$\alpha$-emitting 
galaxies at lower redshifts $z$ $\la$ 0.3 is spatially more extended compared 
to that of the UV continuum. 

The difference between our high-EW(Ly$\alpha$) and high-$z$ galaxies 
may possibly be caused by a low-intensity diffuse Ly$\alpha$ emission in our 
galaxies extending outside the extraction aperture used to obtain the 
one-dimensional spectrum.
In this case, an aperture correction is needed for comparison of the
Ly$\alpha$ flux with fluxes of other hydrogen emission lines.
However, we do not expect a large aperture correction for $L$(Ly$\alpha$)
($>$ 2) because the adopted aperture for the extraction of one-dimensional
spectra includes nearly all light from our galaxies. 
Additionally, we have compared Ly$\alpha$ fluxes in one-dimensional spectra obtained 
with different extraction apertures and reduced with the same 
custom and pipeline data reduction packages. We find that they differ by not
more than 5 per cent. 
Ly$\alpha$ imaging of the lowest-mass SFGs, with stellar masses 
$\la$ 10$^{8}$ M$_\odot$, is needed to verify the importance of aperture 
corrections.

Other factors may play a role in reducing $L$(Ly$\alpha$) in our 
low-mass galaxies, e.g. unaccounted contribution of underlying Ly$\alpha$ 
absorption, enhanced UV extinction due to a larger fraction of dust and/or 
a steeper reddening law in the UV range than in LyC leakers and high-$z$ galaxies.
But these factors are unlikely to be enough to explain the observed large
offset of our galaxies from the extrapolated value (dashed line in 
Fig.~\ref{fig7}a). We note that the H$\beta$ luminosities in our galaxies are
more than one order of magnitude smaller than in the LyC leakers, corresponding
to a smaller number of massive stars by the same factor. Probably, 
dynamical processes such as large-scale outflows due to the evolution of massive
stars in our galaxies are considerably less energetic and may result
in a reduced escaping Ly$\alpha$ emission.

The distribution of the Ly$\alpha$ rest-frame equivalent widths EW(Ly$\alpha$)
for the same samples of galaxies as in Fig.~\ref{fig7}a is shown in 
Fig.~\ref{fig7}b. The range of EW(Ly$\alpha$) for our objects is larger than 
that for LyC leakers with four galaxies having EW(Ly$\alpha$)
smaller than those for LyC leakers, and four galaxies with EW(Ly$\alpha$) 
greater than the average value for LyC leakers. On the other hand, GPs from the 
literature have in general considerably lower EW(Ly$\alpha$). Comparing to 
high-$z$ galaxies we note that the EW(Ly$\alpha$)s of our galaxies are 
several times lower than of high-$z$ galaxies, due to the same reasons (i.e 
possible aperture effects) as for Ly$\alpha$ luminosities.

\citet{Ha17} found a trend of increasing
EW(Ly$\alpha$) with decreasing UV luminosity. They attributed this trend to 
the lower metallicities of lower-luminosity galaxies. However, the distribution of our
galaxies which extend the EW(Ly$\alpha$)-$M_{\rm FUV}$ relation to lower UV luminosities does not support the existence of this 
trend. Although a dependence of EW(Ly$\alpha$) on metallicity is not excluded,
the relation in Fig.~\ref{fig7}b can also be explained by increasingly higher 
EW(Ly$\alpha$) errors at their high values for high-$z$ galaxies and by 
the fact that only galaxies with high EW(Ly$\alpha$) at high redshifts can be
selected. The presence of many low-$z$ GPs with low EW(Ly$\alpha$) favours this 
selection effect.

\subsection{The Ly$\alpha$ escape fraction}

An important quantity to consider is the escape fraction $f_{\rm esc}$(Ly$\alpha$).
In this paper we define $f_{\rm esc}$(Ly$\alpha$) as the ratio of the 
extinction-corrected Ly$\alpha$ to H$\beta$ flux ratio to its Case B value
of 23.3, corresponding to that at a low electron number density, 
$N_{\rm e}$ $\sim$ 10$^2$ cm$^{-3}$ \citep{SH95}:
\begin{equation}
 f_{\rm esc}({\rm Ly}\alpha) = 
\frac{1}{23.3}\frac{I({\rm Ly}\alpha)}{I({\rm H}\beta)}, \label{fesc}
\end{equation}
where $I$(Ly$\alpha$) is the flux density corrected for the Milky Way 
extinction, and $I$(H$\beta$) is the flux density
corrected for both the Milky Way and internal galaxy extinction.
At higher electron densities, for example at $N_{\rm e}$ $\sim$ 10$^3$ cm$^{-3}$, 
the Case B Ly$\alpha$/H$\beta$ ratio is higher by 
$\sim$ 10 per cent, resulting in a lower
$f_{\rm esc}$(Ly$\alpha$).
Using Eq. \ref{fesc}, we calculate $f_{\rm esc}$(Ly$\alpha$) for the eleven LyC leakers
from \citet{I16,I16b,I18,I18b} and compare them with
their directly measured $f_{\rm esc}$(LyC) in Fig.~\ref{fig8}. 
It is seen that there is a general trend of increasing 
$f_{\rm esc}$(LyC) with increasing $f_{\rm esc}$(Ly$\alpha$). That trend 
can be approximated by the maximum likelihood relation (solid line)
\begin{equation}
\log f_{\rm esc}({\rm LyC})=(2.67\pm 0.51)\times \log f_{\rm esc}({\rm Ly}\alpha) 
+ (0.40\pm 0.29), 
\label{LyC-Lya}
\end{equation}
but with large uncertainties.
The distribution of the data points in Fig.~\ref{fig8} is somewhat different from that of the 
dusty, clumpy interstellar medium models of  \citet{D16} (shaded region in
Fig.~\ref{fig8}). Their models predict 
$f_{\rm esc}$(Ly$\alpha$) $\ga$ 2$\times$$f_{\rm esc}$(LyC) and a wide range of 
$f_{\rm esc}$(Ly$\alpha$) at fixed $f_{\rm esc}$(LyC), depending on the covering
factor of the neutral gas. The observational data show a  
much narrower range of $f_{\rm esc}$(Ly$\alpha$) at fixed $f_{\rm esc}$(LyC), with some 
galaxies lying outside the model prediction regions.
A relatively small dispersion of the data points in the observed relation
in Fig.~\ref{fig8} gives hope that 
$f_{\rm esc}$(Ly$\alpha$) can be used to indirectly derive 
$f_{\rm esc}$(LyC). However, the number of confirmed LyC leakers is still 
very small and larger statistics are needed to verify the relation given by 
Eq. \ref{LyC-Lya}.
There is also the possibility that there exist sources with significant 
Ly$\alpha$ escape but with no leaking LyC radiation which would flatten the 
relation shown by the solid line.

The filled circles in Fig.~\ref{fig8} show the galaxies from our sample with 
the exclusion of the 
three galaxies with Ly$\alpha$ absorption. For these  
galaxies, $f_{\rm esc}$(LyC) is derived using the velocity separation between 
the peaks of Ly$\alpha$ emission lines. It is seen that, for a given 
$f_{\rm esc}$(LyC), the galaxies are offset from
the confirmed LyC leakers to lower $f_{\rm esc}$(Ly$\alpha$). Possible aperture 
corrections for Ly$\alpha$ emission would be too small to account for those 
offsets. On the other hand, corrections for unaccounted Ly$\alpha$ 
absorption would increase $f_{\rm esc}$(Ly$\alpha$). Overall, these 
corrections do not appear to be enough to account for the offsets. 
Alternatively, parts of the offsets could be due to an overestimation of the 
indirect value of $f_{\rm esc}$(LyC), as determined by eq. 2 of \citet{I18b}.

It was suggested by \citet{NO14} that the fraction of escaping ionizing 
radiation from galaxies correlates with the O$_{32}$ ratio. However,
\citet{I18b} found that the O$_{32}$ -- $f_{\rm esc}$(LyC) correlation
for the confirmed LyC leakers is rather weak and a high O$_{32}$ does not
garantee a high $f_{\rm esc}$(LyC). Adopting the correlation between the
 Ly$\alpha$ and LyC escape fractions (Fig.~\ref{fig8}), we investigate whether
O$_{32}$ can be used as indirect indicator of escaping ionizing radiation
for galaxies with the highest O$_{32}$. We show in Fig.~\ref{fig9}a the
O$_{32}$ -- $f_{\rm esc}$(Ly$\alpha$) diagram for the same galaxies as in
Fig.~\ref{fig7}, excluding the high-$z$ LAEs. The galaxies
discussed in this paper (filled circles) extend the range of O$_{32}$ up 
to $\sim$ 40, a factor of two higher than in 
previous studies. However, no correlation
is found. Thus, we confirm and strengthen the conclusion of \citet{I18b} that
O$_{32}$ is not a reliable indicator of escaping Ly$\alpha$ emission and likely 
also of escaping LyC radiation. This conclusion is in agreement
with the models of \citet{B19} who have shown that galaxies with various
metallicities and ionization parameters fill all the 
$f_{\rm esc}$(LyC) -- O$_{32}$
space. However, we note that $f_{\rm esc}$(Ly$\alpha$) for our galaxies
(filled circles in Fig.~\ref{fig9}) could be somewhat higher because of 
unaccounted possible aperture corrections for Ly$\alpha$ emission and underlying
absorption.

We discuss other possible indicators of $f_{\rm esc}$(Ly$\alpha$). 
A possible candidate is the rest-frame equivalent width EW(Ly$\alpha$). 
The diagram EW(Ly$\alpha$) -- $f_{\rm esc}$(Ly$\alpha$) is shown in
Fig.~\ref{fig9}b. This diagram is somewhat puzzling. At low EW(Ly$\alpha$),
the Ly$\alpha$ escape fraction $f_{\rm esc}$(Ly$\alpha$) increases
with increasing EW(Ly$\alpha$). However, no such 
dependence is found at high EW(Ly$\alpha$) $\ga$ 60\AA\ for four of our 
galaxies and three confirmed LyC leakers with the highest EW(Ly$\alpha$). 
One possible explanation for that behavior has been proposed by \citet{SM19}. 
They have shown that
the EW(Ly$\alpha$) -- $f_{\rm esc}$(Ly$\alpha$) relation flattens considerably
in the presence of dust absorption and is nearly independent of 
EW(Ly$\alpha$), given a varying ionizing photon production efficiency 
$\xi_{\rm ion}$. In particular, our data, in the flat part of the diagram at 
EW(Ly$\alpha$) $\ga$ 150\AA, can be
reproduced adopting $E(B-V)$ $\sim$ 0.2 mag, corresponding to
$C$(H$\beta$) $\sim$ 0.3, given that log $\xi_{\rm ion}$ increases from
$\sim$ 25.0 to $\sim$ 25.7 with increasing EW(Ly$\alpha$)
(solid line in Fig.~\ref{fig9}b). 
This value of $C$(H$\beta$) is somewhat higher than
the values derived from the SDSS spectra (Table~\ref{tab5}), but is consistent
with the values obtained from the high signal-to-noise ratio LBT spectra for 
J0159$+$0751, J1032$+$4919, J1205$+$4551 and J1355$+$4651 \citep{I17}.

Another indirect indicator of escaping radiation is the velocity separation
between Ly$\alpha$ peaks $V_{\rm sep}$. \citet{I18b} have found a tight 
correlation between $f_{\rm esc}$(LyC) and $V_{\rm sep}$. Fig. \ref{fig9}c
shows a possible correlation between the peak separation $V_{\rm sep}$
and the Ly$\alpha$ escape fraction $f_{\rm esc}$(Ly$\alpha$), described by the 
maximum likelihood relation (solid line in Fig.~\ref{fig9}c):
\begin{equation}
f_{\rm esc}({\rm Ly}\alpha)=6.56\times 10^{-7} V_{\rm sep}^2
-1.25\times 10^{-3}V_{\rm sep}+0.607. \label{fesc1}
\end{equation}
The large scatter of the data around the relation found between an integrated property of Ly$\alpha$
($f_{\rm esc}$(Ly$\alpha$)) and a characteristic of the Ly$\alpha$ profile
($V_{\rm sep}$) is probably to be expected, since several physical properties, such as
dust content, H~{\sc i} column density, H~{\sc i} kinematics and 
geometry affect the Ly$\alpha$ radiative transfer \citep[see e.g. ][]{S11,V15}.

Finally, we consider whether He~{\sc i} emission lines in the optical
range can be indicators of Ly$\alpha$ emission. \citet{I17}
have argued that He~{\sc i} emission line ratios are promising diagnostics 
because some of these lines can be optically thick, indicating a high column
density of the neutral gas. The He~{\sc i} $\lambda$3889\AA\ emission line is
most sensitive to this effect.
This line is blended with the hydrogen H8 $\lambda$3889\AA\ emission line,
and its intensity can be obtained by subtracting
the H8 intensity equal to 0.107$\times$$I$(H$\beta$).

The recombination intensity of the He~{\sc i} $\lambda$3889\AA\ emission line is
$\sim$ 0.1$\times$$I$(H$\beta$). This line can be enhanced by collisions with 
electrons but weakened if its optical depth is high. Therefore, the condition
$I$(He~{\sc i} $\lambda$3889)/$I$(H$\beta$) $<$ 0.1, corresponding to
$I$(H8 + He~{\sc i} $\lambda$3889)/$I$(H$\beta$) $<$ 0.2, would be an indication
of high optical depth and high $N$(H~{\sc i}).

It is seen in Table~\ref{taba1} that 
$I$(H8 + He~{\sc i} $\lambda$3889)/$I$(H$\beta$) is considerably lower than
0.2 in only two galaxies, J0926$+$4504
and J1032$+$4919.  Both of these show broad Ly$\alpha$ absorption line in their
spectra (Fig.~\ref{fig6}d,e), indicating high $N$(H~{\sc i}). For the 
remaining galaxies, 
this technique predicts lower $N$(H~{\sc i}) and thus stronger Ly$\alpha$ in 
emission. The technique fails for only one galaxy, J0007$+$0226, which shows  
Ly$\alpha$ absorption in its spectrum (Fig.~\ref{fig6}a), but has 
$I$(H8 + He~{\sc i} $\lambda$3889)/$I$(H$\beta$) $\sim$0.2. Thus we conclude 
that the He~{\sc i} $\lambda$3889\AA\ emission line is generally a good 
indicator of $N$(H~{\sc i}), and hence of Ly$\alpha$ emission.

\section{Conclusions}\label{summary}

We present here new {\sl HST}/COS observations of eight 
low-redshift ($z$ $<$ 0.07) compact star-forming galaxies (SFG) with 
extremely high O$_{32}$ in the range $\sim$ 22 -- 39.
All studied objects are compact low-mass (log~($M_\star$/M$_\odot$) = 5.8 -- 8.6)
and low-metallicity (12$+$log(O/H)= 7.43 -- 7.82) galaxies with strong nebular
emission lines in their spectra (EW(H$\beta$) = 258 - 483\AA), indicating very 
young starburst ages of 0.5 -- 3.3 Myr. The Ne/O, S/O and Ar/O 
abundance ratios in our galaxies are similar to the values for the bulk
of compact SDSS SFGs. On the other hand, similarly to confirmed LyC leakers,
the N/O abundance ratios in our galaxies are higher than those of the most
compact SDSS SFGs.
We also study the Ly$\alpha$ emission and indirect indicators
of the escaping Ly$\alpha$ and Lyman continuum (LyC) radiation of these SFGs. 
Our main results are summarized as follows:

1. A strong Ly$\alpha$ emission line with two peaks was observed in the
spectra of five galaxies, while a Ly$\alpha$ emission line with a low equivalent
width on top of a broad Ly$\alpha$ absorption profile is present in
the remaining three galaxies. Using the damped wings of these absorption 
profiles, we derive neutral hydrogen column densities $N$(H~{\sc i}) in the 
range (1~--~3)$\times$10$^{21}$ cm$^{-2}$. 

2. We discuss various indirect indicators of escaping Ly$\alpha$ and
ionizing radiation,
such as the O$_{32}$ ratio, EW(Ly$\alpha$) and the velocity separation between 
Ly$\alpha$
emission line peaks $V_{\rm sep}$, assuming that the strength of the Ly$\alpha$
emission line is a measure of LyC leakage. We found that there is no 
correlation between O$_{32}$ and the Ly$\alpha$ escape fraction 
$f_{\rm esc}$(Ly$\alpha$). 
The dependence of $f_{\rm esc}$(Ly$\alpha$) on EW(Ly$\alpha$) is such that
at low EW(Ly$\alpha$) $\la$ 100\AA, there is a linear increase of 
$f_{\rm esc}$(Ly$\alpha$), while at high EW(Ly$\alpha$) $>$ 150\AA, 
$f_{\rm esc}$(Ly$\alpha$) is nearly constant with a value $\sim$ 0.25. This 
behavior may be explained by dust absorption. We find a trend of 
increasing of $f_{\rm esc}$(Ly$\alpha$) with decreasing of $V_{\rm sep}$. However,
even galaxies showing double-peaked Ly$\alpha$ profiles exhibit 
a large scatter between the Ly$\alpha$ escape fraction and the separation of 
their peaks $V_{\rm sep}$. The latter quantity has been shown to correlate more
strongly with the LyC escape fraction \citep{I18b}.

3. Bright compact star-forming regions are seen
in the COS near ultraviolet (NUV) acquisition images. The surface brightness 
profile at the outskirts can be approximated by an exponential disc profile, 
with a scale length of $\sim$ 0.09 -- 0.48 kpc. These scale lengths are several 
times lower than those of confirmed LyC leakers and are among the lowest ones 
found for local blue compact dwarf galaxies.

The global properties of the selected galaxies are very similar 
to those of the lowest-mass high-$z$ galaxies. They are thus ideal nearby 
laboratories for investigating the mechanisms responsible for the
escape of Ly$\alpha$ and ionizing radiation from galaxies during the epoch
of the reionization of the Universe.

\section*{Acknowledgements}

We thank the anonymous referee for valuable comments.
These results are based on observations made with the NASA/ESA 
{\sl Hubble Space Telescope}, 
obtained from the data archive at the Space Telescope Science Institute. 
STScI is operated by the Association of Universities for Research in Astronomy,
Inc. under NASA contract NAS 5-26555. Support for this work was provided by 
NASA through grant number HST-GO-15136.002-A from the Space Telescope Science 
Institute, which is operated by AURA, Inc., under NASA contract NAS 5-26555.
Y.I.I. and N.G.G. acknowledge support from the
National Academy of Sciences of Ukraine (Project No. 0116U003191) and by
its Program of Fundamental Research of the Department of Physics and Astronomy
(Project No. 0117U000240) of the National Academy of Sciences of
Ukraine. G.W. has been supported by the Deutsches Zentrum f\"ur Luft- und 
Raumfahrt (DLR) through grant number 50OR1720.
I.O. acknowledges grants GACR 14-20666P and 17-06217Y of the Czech National 
Foundation. 
Funding for the Sloan Digital Sky Survey IV has been provided by the 
Alfred P. Sloan Foundation, the U.S. Department of Energy Office of Science, 
and the Participating Institutions. SDSS-IV acknowledges
support and resources from the Center for High-Performance Computing at
the University of Utah. The SDSS web site is www.sdss.org.
SDSS-IV is managed by the Astrophysical Research Consortium for the 
Participating Institutions of the SDSS Collaboration.
GALEX is a NASA mission managed by the Jet Propulsion Laboratory.
This publication makes use of data products from the Wide-field Infrared Survey 
Explorer, which is a joint project of the University of California, Los Angeles,
and the Jet Propulsion Laboratory/California Institute of Technology, funded by 
the National Aeronautics and Space AdministrationGALEX is a NASA mission  
managed  by  the  Jet  Propulsion  Laboratory.
This research has made use of the NASA/IPAC Extragalactic Database (NED) which 
is operated by the Jet  Propulsion  Laboratory,  California  Institute  of  
Technology,  under  contract with the National Aeronautics and Space 
Administration.





\begin{thebibliography}{}

\bibitem[Abolfathi et al.(2018)]{A18} Abolfathi B. et al., 2018, \apjs, 235, 42

\bibitem[Ade et al.(2014)]{P14} Ade P. A. R. et al.,
2014, \aap, 571, 16





\bibitem[Baldwin et al.(1981)Baldwin, Phillips \& Terlevich]{BPT81} 
Baldwin J. A., Phillips M. M., Terlevich R., 1981, \pasp, 93, 5

\bibitem[Bassett et al.(2019)]{B19} Bassett et al., 2019, \mnras, 493, 5223

\bibitem[Bian et al.(2017)]{B17} Bian F., Fan X., McGreer I., Cai Z., Jiang L.,
2017, \apj, 837, 12

\bibitem[Bohlin(1975)]{B75} Bohlin R. C., 1975, \apj, 200, 402

\bibitem[Borthakur et al.(2014)]{B14} Borthakur S., Heckman T. M., 
Leitherer C., Overzier R. A., 2014, \sci, 346, 216



\bibitem[Bouwens et al.(2015)]{B15a} Bouwens R. J., Illingworth G. D., 
Oesch P. A., Caruana J., Holwerda B., Smit R.,  Wilkins S., 
2015, \apj, 811, 140

\bibitem[Bouwens et al.(2017a)]{Bo17}  Bouwens R. J., Illingworth G. D., 
Oesch P. A., Atek H, Lam D, Stefanon M., 2017a, \apj, 843, 41

\bibitem[Bouwens et al.(2017b)]{Bo17b} Bouwens R. J., Oesch P. A., 
Illingworth G. D., Ellis R. S., Stefanon M., 2017b, \apj, 843, 129





\bibitem[Cardamone et al.(2009)]{Ca09} Cardamone C. et al.,
2009, \mnras, 399, 1191

\bibitem[Cardelli et al.(1989)Cardelli, Clayton \& Mathis]{C89} 
Cardelli J. A., Clayton G. C., Mathis J. S., 1989, \apj, 345, 245

\bibitem[Chisholm et al.(2017)]{C17} Chisholm J., Orlitov\'a I., 
Schaerer D., Verhamme A., Worseck G., Izotov Y. I., Thuan T. X., Guseva N. G.,
2017, \aap, 605, 67

\bibitem[Chisholm et al.(2018)]{C18} Chisholm J. et al., 2018, \aap, 616, 30


\bibitem[Curtis-Lake et al.(2016)]{CL16} Curtis-Lake E. et al., 
2016, \mnras, 457, 440

\bibitem[de Barros et al.(2016)]{B16} de Barros S. et al., 2016, \aap, 585, 51


\bibitem[Dijkstra et al.(2016)Dijkstra, Gronke \& Venkatesan]{D16} 
Dijkstra M., Gronke M., Venkatesan A., 2016, \apj, 828, 71

\bibitem[Dressler et al.(2015)]{D15} Dressler A., Henry A., Martin C. L., 
Sawicki M., McCarthy P., Villaneuva E., 2015, \apj, 806, 19

\bibitem[Erb et al.(2018)Erb, Steidel \& Chen]{E18} Erb D. K., Steidel C. C.,
Chen Y., 2018, \apjl, 862, 10

\bibitem[Erb et al.(2019)]{E19} Erb D. K., Berg D. A., Auger M. W., 
Kaplan D. L., Brammer G., Pettini M., 2019, \apj, 884, 7


\bibitem[Faisst(2016)]{F16} Faisst A. L., 2016, \apj, 829, 99









\bibitem[Fletcher et al.(2019)]{F19} Fletcher T. J., Tang M., Robertson B. E., 
Nakajima K., Ellis R. S., Stark D. P., Inoue A., 2019, \apj, 878, 87








\bibitem[Grazian et al.(2016)]{Gr15} Grazian A. et al., 2016, \aap, 585, 48 







\bibitem[Guseva et al.(2013)]{G13} Guseva N. G., Izotov Y. I., Fricke K. J.,
Henkel C., 2013, \aap, 555, 90




\bibitem[Harikane et al.(2018)]{Ha18} Harikane Y. et al., 2018, \apj, 859, 84

\bibitem[Hashimoto et al.(2017)]{Ha17} Hashimoto T. et al., 2017, \aap, 608, 10

\bibitem[Hassan et al.(2018)]{H18} Hassan S., Dav\'e R., Mitra S., 
Finlator K., Ciardi B., Santos M. G., 2018, \mnras, 473, 227





\bibitem[Henry et al.(2015)]{H15} Henry A., Scarlata C., Martin C. S., Erb D.,
2015, \apj, 809, 19


\bibitem[Hernandez et al.(2018)]{He18} Hernandez S., Leitherer C., 
Boquien M., Buat V., Burgarella D., Calzetti D., Noll S., 2018,
\mnras, 478, 1292




\bibitem[Inoue et al.(2014)]{Inoue14} Inoue A.~K., Shimizu 
I., Iwata I., Tanaka M., 2014, \mnras, 442, 1805 





\bibitem[Izotov et al.(1994)Izotov, Thuan \& Lipovetsky]{ITL94} Izotov Y. I.,
Thuan T. X., Lipovetsky V. A., 1994, \apj, 435, 647

\bibitem[Izotov et al.(2006)]{I06} Izotov Y. I., Stasi\'nska G., Meynet G.,
Guseva N. G., Thuan T. X., 2006, \aap, 448, 955

\bibitem[Izotov et al.(2011)Izotov, Guseva \& Thuan]{I11} Izotov Y. I., 
Guseva N. G., Thuan T. X., 2011, \apj, 728, 161





\bibitem[Izotov et al.(2016a)]{I16} Izotov Y. I., Orlitov\'a I., Schaerer D.,
Thuan T. X., Verhamme A., Guseva N. G.,  Worseck G., 2016a, \nat, 529, 178

\bibitem[Izotov et al.(2016b)]{I16b} Izotov Y. I., Schaerer D., Thuan, T. X., 
Worseck G., Guseva N. G., Orlitov\'a I., Verhamme A., 2016b, \mnras, 461, 3683

\bibitem[Izotov et al.(2016c)]{I16c} Izotov Y. I., Guseva N. G., 
Fricke K. J.,  Henkel C., 2016c, \mnras, 462, 4427

\bibitem[Izotov et al.(2017)Izotov, Thuan \& Guseva]{I17} Izotov Y. I., 
Thuan T. X., Guseva N. G., 2017, \mnras, 471, 548

\bibitem[Izotov et al.(2018a)]{I18} Izotov Y. I., Schaerer D.,
Worseck G., Guseva N. G., Thuan, T. X., Verhamme A., Orlitov\'a I., Fricke K. J,
 2018a, \mnras, 474, 4514

\bibitem[Izotov et al.(2018b)]{I18b} Izotov Y. I., Worseck G., Schaerer D.,
Guseva N. G., Thuan, T. X., Fricke K. J, Verhamme A., Orlitov\'a I.,
 2018b, \mnras, 478, 4851

\bibitem[Jaskot \& Oey(2013)]{JO13} Jaskot A. E., Oey M. S.,
2013, \apj, 766, 91

\bibitem[Jaskot \& Oey(2014)]{JO14} Jaskot A. E., Oey M. S.,
2014, \apj, 791, L19

\bibitem[Jaskot et al.(2017)]{J17} Jaskot A. E., Oey M. S., Scarlata C.,
Dowd T., 2017, \apj, 851, L9

\bibitem[Jiang et al.(2013)]{J13} Jiang L. et al., 
2018, \apj, 772, 99


\bibitem[Kashikawa et al.(2011)]{K11} Kashikawa N. et al., 2011, \apj, 734, 119

\bibitem[Kauffmann et al.(2003)]{K03} Kauffmann G. et al.,
2003, \mnras, 341, 33

\bibitem[Kennicutt(1998)]{K98} Kennicutt R. C., Jr.,
1998, \araa, 36, 189


\bibitem[Khaire et al.(2016)]{K16} Khaire V., Srianand R., Choudhury T. R.,
Gaikwad P., 2016, \mnras, 457, 4051


\bibitem[Kimm et al.(2019)]{K19} Kimm T., Blaizot J., Carel T., 
Michel-Danzac L., Katz H., Rosdahl J., Verhamme A., Haehnelt M., 2019, \mnras,
486, 2215


\bibitem[Kulkarni et al.(2019)Kulkarni, Worseck \& Hennawi]{Ku19} 
Kulkarni G., Worseck G., Hennawi J. F., 2019, \mnras, 488, 1035


\bibitem[de La Vieuville et al.(2019)]{L19} de La Vieuville G. et al.,
2019, \aap, 628, 3







\bibitem[Leitherer et al.(2016)]{L16} Leitherer C., Hernandez S., 
Lee J. C., Oey M. S., 2016, \apj, 823, L64


\bibitem[Livermore et al.(2017)Livermore, Finkelstein \& Lotz]{L17}
Livermore R. C., Finkelstein S. L., Lotz, J. M., 2017, \apj, 835, 113 

\bibitem[Lodders(2003)]{L03} Lodders K., 2003, \apj, 591, 1220


\bibitem[Madau \& Haardt(2015)]{Madau15} Madau P., Haardt F., 
2015, \apjl, 813, L8

\bibitem[Maseda et al.(2018)]{Ma18} Maseda M. V. et al., 2018, \apj, 865, L1

\bibitem[Matsuoka et al.(2018)]{Mat18} Matsuoka Y. et al., 2018, \apj, 869, 150



\bibitem[McKinney et al.(2019)]{MK19} McKinney J. H., Jaskot A. E., Oey M. S.,
Yun M. S., Dowd T., Lowenthal J. D., 2019, \apj, 874, 52


\bibitem[Mitra et al.(2013)Mitra, Ferrara \& Choudhury]{M13} 
Mitra S., Ferrara A., Choudhury T. R., 2013, \mnras, 428, L1

\bibitem[Mitra et al.(2018)Mitra, Choudhury \& Ferrara]{M18} 
Mitra S.,  Choudhury T. R., Ferrara A., 2018, \mnras, 473, 1416




\bibitem[Naidu et al.(2018)]{N18} Naidu R. P., Forrest B., Oesch P. A., 
Tran K.-V. H., Holden B. P., 2018, \mnras, 478, 791

\bibitem[Nakajima \& Ouchi(2014)]{NO14} Nakajima K.,  Ouchi M.,
2014, \mnras, 442, 900

\bibitem[Nakajima et al.(2018)]{Na18} Nakajima K., Fletcher T., Ellis R. S.,
Robertson B. E., Iwata I., 2018, \mnras, 477, 2098





\bibitem[Orlitov\'a et al.(2018)]{O18} Orlitov\'a I., Verhamme A., Henry A., 
Scarlata C., Jaskot A., Oey M. S., Schaerer D., 2018, \aap, 616, 60


\bibitem[Ouchi et al.(2008)]{O08} Ouchi M. et al., 2008, \apjs, 176, 301

\bibitem[Ouchi et al.(2009)]{O09} Ouchi M. et al., 2009, \apj, 706, 1136



\bibitem[Papaderos et al.(2002)]{P02} Papaderos P., Izotov Y. I., 
Thuan T. X., Noeske K. G., Fricke K. J., Guseva N. G., Green R. F., 
2002, \aap, 393, 461

\bibitem[Parsa et al.(2018)Parsa, Dunlop \& McLure]{P18} Parsa S., Dunlop J.S.,
McLure R. J., 2018, \mnras, 474, 2904


\bibitem[Paulino-Afonso et al.(2018)]{PA18} Paulino-Afonso A. et al., 2018,
\mnras, 476, 5479





\bibitem[Rivera-Thorsen et al.(2017)]{RT17} Rivera-Thorsen T. E. et al., 
2017, \aap, 608, L4

\bibitem[Rivera-Thorsen et al.(2019)]{RT19} Rivera-Thorsen T. E. et al., 
2019, preprint arXiv:1904.08186

\bibitem[Robertson et al.(2013)]{R13} Robertson B. E. et al.,
2013, \apj, 768, 71

\bibitem[Robertson et al.(2015)]{Robertson15} Robertson B.~E., 
Ellis R.~S., Furlanetto S.~R., Dunlop J.~S., 2015, \apjl, 802, L19 



\bibitem[Rutkowski et al.(2017)]{R17} Rutkowski M. G. et al., 2017, \apj, 
841, L27



\bibitem[Schaerer et al.(2011)]{S11} Schaerer D., Hayes M., Verhamme A.,
Teyssier R., 2011, \aap, 531, 12

\bibitem[Schaerer et al.(2016)]{S16} Schaerer D., Izotov Y. I., Verhamme A.,
Orlitov\'a I., Thuan T. X., Worseck G., Guseva, N. G., 2016, \aap, 591, L8


\bibitem[Shapley et al.(2016)]{Sh16} Shapley A. E., Steidel C. C., 
Strom A. L., Bogosavljevi\'c M., Reddy N. A., Siana B. Mostardi R. E., 
Rudie G. C., 2016, \apj, 826, L24




\bibitem[Smith et al.(2018)]{Sm18} Smith B. M. et al., 2018, \apj, 853, 191

\bibitem[Sobral \& Matthee(2019)]{SM19} Sobral D., Matthee J., 2019, \aap, 623,
157

\bibitem[Stasi\'nska et al.(2015)]{S15} Stasi\'nska G., Izotov Y., 
Morisset C.,  Guseva N., 2015, \aap, 576, 83



\bibitem[Steidel et al.(2018)]{St18} Steidel C. C., Bogosavljevi\'c M., 
Shapley A.E., Reddy N. A., Rudie G. C., Pettini M., Trainor R. F., 
Strom A. L., 2018, \apj, 869, 123

\bibitem[Storey \& Hummer(1995)]{SH95} Storey P. J.,  Hummer D. G.,
1995, \mnras, 272, 41







\bibitem[Vanzella et al.(2010)]{V10} Vanzella E. et al., 
2010, \apj, 725, 1011 


\bibitem[Vanzella et al.(2012)]{V12} Vanzella E. et al., 2012, \apj, 751, 70 

\bibitem[Vanzella et al.(2015)]{Va15} Vanzella E. et al., 2015, \aap, 576, 116 

\bibitem[Vanzella et al.(2018)]{Va18} Vanzella E. et al., 2018, \mnras, 476, 
L15 

\bibitem[Verhamme et al.(2015)]{V15}  Verhamme A., Orlitov\'a I., 
Schaerer D., Hayes M., 2015, \aap, 578, 7

\bibitem[Verhamme et al.(2017)]{V17} Verhamme A., Orlitov\'a I., Schaerer D., 
Izotov Y., Worseck G., Thuan T. X., Guseva N., 2017, \aap, 597, 13



\bibitem[Wise \& Chen(2009)]{WC09} Wise J. H., Cen R.,
2009, \apj, 693, 984


\bibitem[Wisotzki et al.(2016)]{Wi16} Wisotzki L. et al., 2016, \aap, 587, 98

\bibitem[Wisotzki et al.(2018)]{Wi18} Wisotzki L. et al., 2018, \nat, 562, 229


\bibitem[Worseck et al.(2016)]{W16} Worseck G., Prochaska J. X., Hennawi J. F.,
McQuinn M., 2016, \apj, 825, 144

\bibitem[Wright(2006)]{W06} Wright E. L., 2006, \pasp, 118, 1711

\bibitem[Yajima et al.(2011)Yajima, Choi \& Nagamine]{Y11} 
Yajima H., Choi J.-H., Nagamine K., 2011, \mnras, 412, 411


\bibitem[Yang et al.(2017a)]{Y17} Yang H. et al., 2017a, \apj, 844, 171

\bibitem[Yang et al.(2017b)]{Y17b} Yang H., Malhotra S., Rhoads J. E., 
Leitherer C., Wofford A., Jiang T., Wang J., 2017b, \apj, 838, 4


\end{thebibliography}


\appendix

\section{Emission-line fluxes in SDSS spectra and element abundances}

  \begin{table*}
  \caption{Extinction-corrected emission-line fluxes in SDSS spectra
\label{taba1}}
\begin{tabular}{lcrrrrrrrr} \hline
 & &\multicolumn{8}{c}{100$\times$$I$($\lambda$)/$I$(H$\beta$)$^{\rm a}$}\\
Line &\multicolumn{1}{c}{$\lambda$}& 0007$+$0226& 0159$+$0751& 0820$+$5431~\,& 0926$+$4504& 1032$+$4919~\,& 1205$+$4551& 1242$+$4851& 1355$+$4651\\
\hline
$[$O~{\sc ii}$]$     &3727&  24.1$\pm$~\,1.7& 18.4$\pm$~\,1.8& 28.0$\pm$~\,2.8~\,& 22.7$\pm$~\,2.0& 25.0$\pm$~\,1.2~\,& 20.3$\pm$~\,1.2& 20.3$\pm$~\,1.9& 22.8$\pm$~\,1.6\\
H12                  &3750&   2.3$\pm$~\,1.2&  4.6$\pm$~\,1.6&  7.6$\pm$~\,2.0~\,&  5.0$\pm$~\,1.3&  3.9$\pm$~\,2.9~\,&  4.3$\pm$~\,0.7&\multicolumn{1}{c}{...}&\multicolumn{1}{c}{...}\\
H11                  &3771&   4.5$\pm$~\,1.1&  4.7$\pm$~\,1.5&  6.8$\pm$~\,2.6~\,&  5.4$\pm$~\,1.3&  4.6$\pm$~\,2.9~\,&  5.7$\pm$~\,0.8&\multicolumn{1}{c}{...}&  4.8$\pm$~\,1.1\\
H10                  &3798&   4.9$\pm$~\,1.1&  7.2$\pm$~\,1.6&  7.9$\pm$~\,2.2~\,&  6.5$\pm$~\,1.6&  6.1$\pm$~\,2.8~\,&  6.9$\pm$~\,0.8&  6.7$\pm$~\,1.2&  8.8$\pm$~\,1.1\\
H9                   &3836&   7.1$\pm$~\,1.1&  8.6$\pm$~\,1.5& 10.7$\pm$~\,2.2~\,&  6.5$\pm$~\,1.6&  7.9$\pm$~\,3.3~\,&  8.6$\pm$~\,0.8&  9.2$\pm$~\,1.3&  9.9$\pm$~\,1.0\\
$[$Ne~{\sc iii}$]$   &3869&  59.2$\pm$~\,3.1& 58.1$\pm$~\,3.4& 50.1$\pm$~\,3.8~\,& 47.5$\pm$~\,3.1& 48.0$\pm$~\,2.0~\,& 28.2$\pm$~\,1.4& 49.4$\pm$~\,3.1& 53.6$\pm$~\,2.8\\
H8+He~{\sc i}        &3889&  20.2$\pm$~\,1.6& 20.0$\pm$~\,1.9& 21.0$\pm$~\,2.5~\,& 14.6$\pm$~\,2.1& 14.9$\pm$~\,3.4~\,& 17.6$\pm$~\,1.1& 24.2$\pm$~\,4.4& 23.3$\pm$~\,1.6\\
H7+$[$Ne~{\sc iii}$]$&3969&  34.7$\pm$~\,2.1& 34.0$\pm$~\,2.4& 33.8$\pm$~\,3.5~\,& 29.8$\pm$~\,2.5& 32.2$\pm$~\,4.0~\,& 27.2$\pm$~\,1.4& 30.1$\pm$~\,3.4& 38.4$\pm$~\,2.3\\
H$\delta$            &4101&  26.7$\pm$~\,1.8& 26.5$\pm$~\,2.0& 31.7$\pm$~\,3.0~\,& 25.2$\pm$~\,2.2& 25.8$\pm$~\,3.8~\,& 28.4$\pm$~\,1.4& 28.8$\pm$~\,3.4& 26.8$\pm$~\,1.8\\
H$\gamma$            &4340&  47.3$\pm$~\,2.6& 47.5$\pm$~\,2.8& 48.6$\pm$~\,3.5~\,& 44.0$\pm$~\,3.0& 44.8$\pm$~\,4.5~\,& 47.5$\pm$~\,2.0& 49.7$\pm$~\,4.4& 48.3$\pm$~\,2.6\\
$[$O~{\sc iii}$]$    &4363&  19.7$\pm$~\,1.4& 22.2$\pm$~\,1.7& 21.6$\pm$~\,2.3~\,& 19.8$\pm$~\,1.7& 19.8$\pm$~\,1.0~\,& 13.3$\pm$~\,0.8& 19.6$\pm$~\,1.7& 20.7$\pm$~\,1.4\\
He~{\sc i}           &4471&   3.5$\pm$~\,0.8&  4.2$\pm$~\,0.8&  3.0$\pm$~\,1.4~\,&  4.4$\pm$~\,1.0&  4.0$\pm$~\,0.4~\,&  4.3$\pm$~\,0.5&  4.6$\pm$~\,1.1&  3.0$\pm$~\,0.7\\
He~{\sc ii}          &4686&   1.3$\pm$~\,0.6&  1.4$\pm$~\,0.6&\multicolumn{1}{c}{...}&  0.9$\pm$~\,0.7&  1.2$\pm$~\,0.3~\,&  3.8$\pm$~\,0.5&  2.2$\pm$~\,0.9&  2.7$\pm$~\,0.6\\
H$\beta$             &4861& 100.0$\pm$~\,4.5&100.0$\pm$~\,4.8&100.0$\pm$~\,6.0~\,&100.0$\pm$~\,5.2&100.0$\pm$~\,5.4~\,&100.0$\pm$~\,3.7&100.0$\pm$~\,3.7&100.0$\pm$~\,4.5\\
$[$O~{\sc iii}$]$    &4959& 260.0$\pm$10.5  &213.2$\pm$~\,9.1&197.3$\pm$10.5~\,  &223.6$\pm$10.3  &209.8$\pm$~\,7.1~\,&136.2$\pm$~\,4.9&174.7$\pm$~\,8.0&197.2$\pm$~\,8.1\\
$[$O~{\sc iii}$]$    &5007& 775.0$\pm$29.9  &630.6$\pm$25.6  &568.2$\pm$24.3~\,  &674.0$\pm$29.1  &  4.3$\pm$~\,0.4$^{\rm b}$&413.1$\pm$14.1  &501.9$\pm$20.1  &599.4$\pm$23.2  \\
He~{\sc i}           &5876&  10.7$\pm$~\,1.0& 11.3$\pm$~\,1.0&  7.6$\pm$~\,1.3~\,& 11.9$\pm$~\,1.2& 12.0$\pm$~\,0.7~\,& 12.0$\pm$~\,0.7&  9.9$\pm$~\,1.1&  8.7$\pm$~\,0.9\\
$[$O~{\sc i}$]$      &6300&   0.5$\pm$~\,0.4&  0.5$\pm$~\,0.4&\multicolumn{1}{c}{...}&\multicolumn{1}{c}{...}&  0.9$\pm$~\,0.2~\,&  1.5$\pm$~\,0.3&  0.5$\pm$~\,0.6&  0.9$\pm$~\,0.4\\
$[$S~{\sc iii}$]$    &6312&   0.9$\pm$~\,0.5&  0.7$\pm$~\,0.4&\multicolumn{1}{c}{...}&  0.9$\pm$~\,0.5&  1.1$\pm$~\,0.7~\,&  0.6$\pm$~\,0.2&  0.7$\pm$~\,0.6&  1.2$\pm$~\,0.4\\
H$\alpha$            &6563& 277.3$\pm$12.1  &275.3$\pm$12.4  &231.8$\pm$13.1$^{\rm b}$&275.3$\pm$13.5  &132.9$\pm$~\,5.6$^{\rm b}$&274.7$\pm$10.3  &272.0$\pm$14.5  &273.2$\pm$11.8  \\
$[$N~{\sc ii}$]$     &6583&   1.1$\pm$~\,0.5&  1.2$\pm$~\,0.4&\multicolumn{1}{c}{...}&  1.6$\pm$~\,0.6&  2.0$\pm$~\,0.3~\,&  5.1$\pm$~\,0.5&  1.2$\pm$~\,0.5&  0.6$\pm$~\,0.4\\
He~{\sc i}           &6678&   2.9$\pm$~\,0.5&  2.3$\pm$~\,0.5&  2.8$\pm$~\,0.9~\,&  2.9$\pm$~\,0.7&  3.2$\pm$~\,0.4~\,&  2.8$\pm$~\,0.3&  3.2$\pm$~\,0.7&  3.2$\pm$~\,0.5\\
$[$S~{\sc ii}$]$     &6717&   2.9$\pm$~\,0.5&  1.3$\pm$~\,0.4&  3.0$\pm$~\,0.9~\,&  3.2$\pm$~\,0.7&  2.6$\pm$~\,0.3~\,&  1.1$\pm$~\,0.3&  2.4$\pm$~\,0.6&  2.3$\pm$~\,0.4\\
$[$S~{\sc ii}$]$     &6731&   2.2$\pm$~\,0.5&  1.3$\pm$~\,0.4&  1.4$\pm$~\,0.8~\,&  2.5$\pm$~\,0.6&  2.5$\pm$~\,0.3~\,&  1.3$\pm$~\,0.3&  2.5$\pm$~\,0.6&  2.6$\pm$~\,0.5\\
He~{\sc i}           &7065&   3.4$\pm$~\,0.6&  6.7$\pm$~\,0.7&  3.6$\pm$~\,0.9~\,&  7.6$\pm$~\,0.9&  7.7$\pm$~\,0.5~\,&  7.7$\pm$~\,0.5&  3.7$\pm$~\,0.7&  2.8$\pm$~\,0.5\\
$[$Ar~{\sc iii}$]$   &7136&   3.7$\pm$~\,0.6&  2.0$\pm$~\,0.4&  3.2$\pm$~\,0.8~\,&  2.7$\pm$~\,0.6&  3.0$\pm$~\,0.3~\,&  1.8$\pm$~\,0.3&  2.3$\pm$~\,0.6&  2.6$\pm$~\,0.5\\ \\
$C$(H$\beta$)$^{\rm c}$   &&\multicolumn{1}{c}{0.130}&\multicolumn{1}{c}{0.425}&\multicolumn{1}{c}{0.000}&\multicolumn{1}{c}{0.075}&\multicolumn{1}{c}{0.000}&\multicolumn{1}{c}{0.160}&\multicolumn{1}{c}{0.085}&\multicolumn{1}{c}{0.065}\\
EW(H$\beta$)$^{\rm d}$      &&\multicolumn{1}{c}{ 258$\pm$7 }&\multicolumn{1}{c}{ 322$\pm$9 }&\multicolumn{1}{c}{ 335$\pm$12}&\multicolumn{1}{c}{ 328$\pm$10}&\multicolumn{1}{c}{444$\pm$27}&\multicolumn{1}{c}{ 483$\pm$9 }&\multicolumn{1}{c}{ 360$\pm$10}&\multicolumn{1}{c}{ 269$\pm$7}\\
EW($\lambda$5007)$^{\rm d}$ &&\multicolumn{1}{c}{2180$\pm$20}&\multicolumn{1}{c}{1964$\pm$22}&\multicolumn{1}{c}{2200$\pm$45}&\multicolumn{1}{c}{1819$\pm$22}&\multicolumn{1}{c}{ 19$\pm$7$^{\rm b}$}&\multicolumn{1}{c}{1718$\pm$15}&\multicolumn{1}{c}{1531$\pm$35}&\multicolumn{1}{c}{1541$\pm$15  }\\
EW(H$\alpha$)$^{\rm d}$     &&\multicolumn{1}{c}{1131$\pm$19}&\multicolumn{1}{c}{1629$\pm$27}&\multicolumn{1}{c}{1335$\pm$37$^{\rm b}$}&\multicolumn{1}{c}{1733$\pm$35}&\multicolumn{1}{c}{932$\pm$50$^{\rm b}$}&\multicolumn{1}{c}{1981$\pm$23}&\multicolumn{1}{c}{2210$\pm$60}&\multicolumn{1}{c}{1644$\pm$27  }\\
$F$(H$\beta$)$^{\rm e}$     &&\multicolumn{1}{c}{46.6$\pm$1.2}&\multicolumn{1}{c}{37.4$\pm$1.0}&\multicolumn{1}{c}{20.2$\pm$0.7}&\multicolumn{1}{c}{29.0$\pm$0.9}&\multicolumn{1}{c}{109.0$\pm$1.8}&\multicolumn{1}{c}{90.3$\pm$1.6}&\multicolumn{1}{c}{33.0$\pm$1.0}&\multicolumn{1}{c}{44.1$\pm$1.1}\\
\hline
  \end{tabular}

\hbox{$^{\rm a}$$I$($\lambda$) and $I$(H$\beta$) are emission-line
fluxes, corrected for the extinction derived from the Balmer decrement of hydrogen lines.}

\hbox{$^{\rm b}$Clipped line.}

\hbox{$^{\rm c}$$C$(H$\beta$) is the extinction coefficient derived from the Balmer decrement of
hydrogen lines.}



\hbox{$^{\rm d}$Rest-frame equivalent width in \AA.}

\hbox{$^{\rm e}$$F$(H$\beta$) is the observed H$\beta$ flux density in 10$^{-16}$ erg s$^{-1}$ cm$^{-2}$.}

  \end{table*}

  \begin{table*}
  \caption{Electron temperatures, electron number densities and 
element abundances in H~{\sc ii} regions \label{taba2}}
  \begin{tabular}{lcccc} \hline
&\multicolumn{4}{c}{Galaxy} \\ 
Parameter                              &0007$+$0226    &0159$+$0751    &0820$+$5431    &0926$+$4504 \\
\hline
$T_{\rm e}$ ($[$O {\sc iii}$]$), K        & 17090$\pm$700 & 20510$\pm$960 & 21410$\pm$950 & 18470$\pm$970 \\
$T_{\rm e}$ ($[$O {\sc ii}$]$), K         & 14980$\pm$570 & 15630$\pm$740 & 15610$\pm$830 & 15380$\pm$750 \\
$T_{\rm e}$ ($[$S {\sc iii}$]$), K        & 15620$\pm$580 & 18610$\pm$880 & 19110$\pm$890 & 17140$\pm$810 \\
$N_{\rm e}$ ($[$S {\sc ii}$]$), cm$^{-3}$         &    90$\pm$40  &   660$\pm$300 &    10$\pm$10  &   140$\pm$50  \\ 
O$^+$/H$^+$$\times$10$^{5}$            &0.22$\pm$0.03  &0.16$\pm$0.02  &0.22$\pm$0.04  &0.19$\pm$0.03  \\
O$^{2+}$/H$^+$$\times$10$^{5}$         &6.23$\pm$0.65  &3.43$\pm$0.41  &2.86$\pm$0.46  &4.55$\pm$0.58  \\
O$^{3+}$/H$^+$$\times$10$^{6}$         &0.88$\pm$0.44  &0.66$\pm$0.34  &     ...       &0.45$\pm$0.37  \\
O/H$\times$10$^{5}$                    &6.54$\pm$0.65  &3.66$\pm$0.42  &3.08$\pm$0.47  &4.79$\pm$0.58 \\
12+log(O/H)                             &7.82$\pm$0.04  &7.56$\pm$0.05  &7.49$\pm$0.07  &7.68$\pm$0.05 \\ 
N$^+$/H$^+$$\times$10$^{7}$            &0.81$\pm$0.26  &0.84$\pm$0.22  &     ...       &1.11$\pm$0.33  \\
ICF(N)$^{\rm a}$                       &25.01          &20.48          &     ...       &21.61          \\
N/H$\times$10$^{6}$                    &2.02$\pm$0.78  &1.71$\pm$0.50  &     ...       &2.40$\pm$0.81 \\
log(N/O)                                &~$-$1.51$\pm$0.17~~\,&~$-$1.33$\pm$0.14~~\,&    ...          &~$-$1.30$\pm$0.16~~\,\\ 
Ne$^{2+}$/H$^+$$\times$10$^{5}$        &1.10$\pm$0.12  &0.69$\pm$0.08  &0.54$\pm$0.09  &0.73$\pm$0.10  \\
ICF(Ne)$^{\rm a}$                      &1.00           &1.02           &1.03           &1.01           \\
Ne/H$\times$10$^{5}$                   &1.10$\pm$0.13  &0.71$\pm$0.09  &0.56$\pm$0.10  &0.74$\pm$0.10 \\
log(Ne/O)                               &~$-$0.77$\pm$0.07~~\,&~$-$0.71$\pm$0.07~~\,&~$-$0.74$\pm$0.10~~\,&~$-$0.81$\pm$0.08~~\,\\ 
S$^{+}$/H$^+$$\times$10$^{7}$          &0.49$\pm$0.08  &0.24$\pm$0.09  &     ...       &0.54$\pm$0.10  \\
S$^{2+}$/H$^+$$\times$10$^{6}$         &0.40$\pm$0.21  &0.20$\pm$0.11  &     ...       &0.32$\pm$0.19  \\
ICF(S)$^{\rm a}$                       &3.21           &2.56           &     ...       &2.53           \\
S/H$\times$10$^{6}$                    &1.44$\pm$0.68  &0.57$\pm$0.28  &     ...       &0.94$\pm$0.49 \\
log(S/O)                                &~$-$1.66$\pm$0.21~~\,&~$-$1.81$\pm$0.22~~\,&     ...          &~$-$1.71$\pm$0.23~~\,\\ 
Ar$^{2+}$/H$^+$$\times$10$^{7}$        &1.39$\pm$0.22  &0.56$\pm$0.13  &0.86$\pm$0.23  &0.87$\pm$0.20  \\
ICF(Ar)$^{\rm a}$                      &2.28           &2.03           &1.56           &2.09           \\
Ar/H$\times$10$^{7}$                   &3.16$\pm$1.34  &1.14$\pm$0.81  &1.34$\pm$0.09  &1.82$\pm$1.21 \\
log(Ar/O)                               &~$-$2.32$\pm$0.19~~\,&~$-$2.51$\pm$0.31~~\,&~$-$2.36$\pm$0.31~~\,&~$-$2.42$\pm$0.29~~\,\\ 
\hline
&\multicolumn{4}{c}{Galaxy} \\ 
Parameter                              &1032$+$4919    &1205$+$4551    &1242$+$4851    &1355$+$4651\\ 
\hline
$T_{\rm e}$ ($[$O {\sc iii}$]$), K        & 19540$\pm$620 & 19500$\pm$790 & 21750$\pm$930 & 20340$\pm$950 \\
$T_{\rm e}$ ($[$O {\sc ii}$]$), K         & 15570$\pm$460 & 15560$\pm$590 & 15590$\pm$870 & 15630$\pm$670  \\
$T_{\rm e}$ ($[$S {\sc iii}$]$), K        & 18090$\pm$510 & 18260$\pm$660 & 19400$\pm$910 & 18560$\pm$790 \\
$N_{\rm e}$ ($[$S {\sc ii}$]$), cm$^{-3}$         &   560$\pm$50  &  1090$\pm$580 &   770$\pm$170 &  1200$\pm$150 \\ 
O$^+$/H$^+$$\times$10$^{5}$            &0.21$\pm$0.02  &0.18$\pm$0.02  &0.18$\pm$0.03  &0.21$\pm$0.03 \\
O$^{2+}$/H$^+$$\times$10$^{5}$         &3.66$\pm$0.28  &2.48$\pm$0.24  &2.45$\pm$0.34  &3.30$\pm$0.36 \\
O$^{3+}$/H$^+$$\times$10$^{6}$         &0.43$\pm$0.12  &1.09$\pm$0.21  &0.55$\pm$0.26  &0.89$\pm$0.27 \\
O/H$\times$10$^{5}$                   &3.92$\pm$0.28  &2.77$\pm$0.24  &2.68$\pm$0.34  &3.59$\pm$0.36 \\
12+log(O/H)                           &7.59$\pm$0.03  &7.44$\pm$0.04  &7.43$\pm$0.06  &7.56$\pm$0.04 \\ 
N$^+$/H$^+$$\times$10$^{7}$           &1.38$\pm$0.17  &3.53$\pm$0.33  &0.81$\pm$0.29  &0.40$\pm$0.19 \\
ICF(N)$^{\rm a}$                       &16.24          &13.50          &13.60          &15.57  \\
N/H$\times$10$^{6}$                   &2.25$\pm$0.31  &4.77$\pm$0.50  &1.10$\pm$0.42  &0.63$\pm$0.34 \\
log(N/O)                               &~$-$1.24$\pm$0.07~~\,&~$-$0.76$\pm$0.06~~\,&~$-$1.39$\pm$0.18~~\,&~$-$1.76$\pm$0.24~~\, \\ 
Ne$^{2+}$/H$^+$$\times$10$^{5}$        &0.64$\pm$0.05  &0.37$\pm$0.04  &0.52$\pm$0.07  &0.65$\pm$0.07 \\
ICF(Ne)$^{\rm a}$                      &1.03           &1.04           &1.04           &1.03  \\
Ne/H$\times$10$^{5}$                  &0.66$\pm$0.05  &0.40$\pm$0.04  &0.54$\pm$0.09  &0.68$\pm$0.08 \\
log(Ne/O)                             &~$-$0.77$\pm$0.05~~\,&~$-$0.85$\pm$0.06~~\,&~$-$0.70$\pm$0.08~~\,&~$-$0.73$\pm$0.07~~\, \\ 
S$^{+}$/H$^+$$\times$10$^{7}$          &0.48$\pm$0.05  &0.24$\pm$0.06  &0.48$\pm$0.13  &0.49$\pm$0.10 \\
S$^{2+}$/H$^+$$\times$10$^{6}$         &0.32$\pm$0.07  &0.17$\pm$0.07  &0.16$\pm$0.14  &0.32$\pm$0.11 \\
ICF(S)$^{\rm a}$                       &2.03           &2.37           &2.45           &2.17  \\
S/H$\times$10$^{6}$                    &0.74$\pm$0.15  &0.47$\pm$0.16  &0.52$\pm$0.35  &0.81$\pm$0.25 \\
log(S/O)                               &~$-$1.72$\pm$0.09~~\,&~$-$1.77$\pm$0.16~~\,&~$-$1.72$\pm$0.30~~\,&~$-$1.65$\pm$0.14~~\, \\ 
Ar$^{2+}$/H$^+$$\times$10$^{7}$        &0.88$\pm$0.10  &0.54$\pm$0.08  &0.62$\pm$0.16  &0.74$\pm$0.14 \\
ICF(Ar)$^{\rm a}$                      &1.78           &1.62           &1.63           &1.74 \\
Ar/H$\times$10$^{7}$                  &1.57$\pm$0.39  &0.87$\pm$0.34  &1.01$\pm$0.70  &1.28$\pm$0.59 \\
log(Ar/O)                             &~$-$2.40$\pm$0.11~~\,&~$-$2.50$\pm$0.17~~\,&~$-$2.43$\pm$0.31~~\,&~$-$2.45$\pm$0.21~~\, \\ 
\hline
\end{tabular}

\hbox{$^{\rm a}$ICF is the ionization correction factor.}
  \end{table*}

\bsp	
\label{lastpage}
\end{document}